\documentstyle[a4,amsmath,amssymb,latexsym,epsfig,citesort]{article}
\def \d{{\mathrm{d}}}

\def \pd{\partial}

\def \e{{\mathrm{e}}}

\def \tl#1{\overset{\kern 1pt\circ}{#1}}
\def \TL#1{\overset{\kern -3pt \circ}{#1}}
\def \TLL#1{\overset{\kern -7pt \circ}{#1}}

\begin{document}
\title{{\bf A nonsingular solution of the edge dislocation in the gauge theory of dislocations}}
\author{Markus Lazar\\
        Max-Planck-Institute for Mathematics in the Sciences,\\
        Inselstr. 22-26, D-04103 Leipzig, Germany\\
        E-mail: lazar@mis.mpg.de}

\date{\today}
%%%%\date{August 5, 2002}    
\maketitle
%%\vspace*{-5mm}
\begin{abstract}
A (linear) nonsingular solution for the edge dislocation in the translational gauge theory 
of defects is presented. 
The stress function method is used and a modified stress function is obtained.
All field quantities are globally defined and the solution agrees with the 
classical solution for the edge dislocation in the far field.
The components of the stress, strain, distortion and displacement field
are also defined in the dislocation core region and they have no singularity
there. 
The dislocation density, moment and couple stress
for an edge dislocation are calculated.
The solutions for the stress and strain field obtained here are in
agreement with those obtained by Gutkin and Aifantis through an 
analysis of the edge dislocation in the strain gradient elasticity. 
Additionally, the relation between the gauge theory and Eringen's so-called 
nonlocal theory of dislocations is given.
%%%\end{abstract}
\\

\noindent
{\bf Keywords:} elastoplasticity, dislocation theory, gauge theory of defects,
strain gradient, nonlocal elasticity \\ 
{\bf PACS:} 61.72.Lk, 62.20.-x, 81.40.Jj
\end{abstract}
\rightline{\tt MPI-MIS 65/2002}
\rightline{\tt cond-mat/0208360}
\vspace*{2mm}
\noindent
 \section{Introduction}
\setcounter{equation}{0}
The gauge theory of 
dislocations~\cite{Lazar00,Lazar02,Lazar02b,Malyshev00,GL79,Edelen82,Edelen83,Edelen88,VS88,Edelen96,Popov,Kadic94,KV92,Gairola81,Gairola93,Kroener93,Kroener96,Kunin85,Kleinert89,Osipov,Volkov,Bogatov} 
is a very attractive field of research. 
The theory of dislocations is in these theories considered as a 
translational gauge theory ($T(3)$-gauge theory). 
An important goal is to find modified
solutions of the stress and strain field of screw and edge dislocations in 
this framework because the ``classical'' elasticity of dislocations
contains singularities at the dislocation line. Therefore, the classical solutions
cannot be applied within the dislocation core region. 
This is unfortunate since the core region is the most important from the
point of dislocation and failure mechanics.
The difference in these gauge models is the form of the $T(3)$-gauge Lagrangian.
Edelen~{\it et al.}~\cite{Edelen82,Edelen83,Edelen88,Edelen96} used the
simplest form quadratic in the $T(3)$-gauge field strength.
In~\cite{Lazar00,Lazar02,Lazar02b,Malyshev00,KV92} the gauge Lagrangian was proposed
in the three-dimensional Hilbert-Einstein form. The most general dislocation 
gauge Lagrangian for an isotropic and anisotropic material was 
proposed in~\cite{Lazar00,Lazar02}. 

A correct gauge theoretical solution of the stress field of a screw 
dislocation is given in~\cite{Lazar02,Lazar02b,Malyshev00,VS88,Edelen96}. 
This solution has no singularity at the dislocation line and the far field 
stress is that of a classical screw dislocation. 

Up to now, the situation is less satisfactory for the edge dislocation.
In the Edelen model~\cite{Edelen82,Edelen83,Edelen88} the far field of the stress 
$\sigma_{zz}$ is different from the standard one. 
In fact, the far field stress $\sigma_{zz}$ is equal to
$\nu^{-1}(\sigma_{xx}+\sigma_{yy})$ instead of 
$\nu(\sigma_{xx}+\sigma_{yy})$ (see also~\cite{Malyshev00}).
Unfortunately, no explicit gauge solution for the edge dislocation was given 
in~\cite{Edelen82,Edelen83,Edelen88}. 
Moreover, the structure of the gauge Lagrangian used by Edelen~{\it et al.} 
requires, in principle, an asymmetric force stress~\cite{Lazar02}.
Malyshev~\cite{Malyshev00} used the Hilbert-Einstein 
Lagrangian as a gauge Lagrangian and found the stress field of 
a straight edge dislocation with a modified asymptotic behaviour 
which differs from the classical edge dislocation. 
In fact, the stress field has no singularity at the dislocation line
and has an oscillatory behaviour with slowly decreasing amplitude far
from the dislocation core. 
His solution can be considered as a complicated gauge defect with 
modified asymptotics instead of an elementary defect~\cite{Malyshev00}.

On the other hand, there are other non-standard continuum models of 
defects, e.g., the Peierls-Nabarro model~\cite{Peierls40,Nabarro47,Nabarro,HL}, 
the nonlocal continuum model~\cite{Eringen77,Eringen83,AE83,Eringen85,Eringen87} 
and the strain gradient elasticity~\cite{AA92,Aifantis94,AA97,GA96,GA97,GA99,Gutkin00}.
The force stresses calculated in these theories are symmetric 
even in the core. 
In the case of a screw dislocation, the solution of the force stresses in nonlocal 
elasticity~\cite{Eringen83,Eringen85,Eringen87} and in gradient elasticity~\cite{GA99,Gutkin00}
agrees with the gauge theoretical one~\cite{Lazar02,Lazar02b,Malyshev00,VS88,Edelen96}. 
Additionally, the strain fields calculated in gradient theory~\cite{GA96,GA99,Gutkin00}
and in gauge theory~\cite{Lazar02,Lazar02b,VS88,Edelen96} coincide.
In the case of an edge dislocation, the solution in nonlocal elasticity
was given in an integral form by means of a special two-dimensional nonlocal
kernel~\cite{Eringen85}. 
Using another two-dimensional nonlocal kernel Eringen~\cite{Eringen77} gave a 
different solution of the straight edge dislocation.
The dislocation stress field 
in gradient elasticity~\cite{GA97,GA99,Gutkin00} was obtained in a closed 
analytical form by using the Fourier transform method, 
it is equal to zero at the dislocation line and beyond the
dislocation core the classical and the gradient solutions coincide. 
In a quasi-continuum model~\cite{Eringen87,Brailsford66,Kunin86,VK93,VK95} 
a modified stress field of an edge and a screw dislocation was obtained, also. 
In this approach the stress field has no divergence at the dislocation line 
and at large distances it contains a decreasing oscillatory contribution.
Dislocations have also been considered in a Cosserat 
media~(see, e.g.,~\cite{Gunter58,Schaefer67,Kluge69}). 
The force stresses of dislocations calculated within the Cosserat theory 
are, in general, asymmetric.
The solutions of edge and screw 
dislocations~\cite{Misicu65,Teodosiu65,Kessel70,Knesl72,Nowacki73,Nowacki74,Minagawa77}
found for the elastic strain and stress fields differ from the classical solutions 
but still posses singularities at the dislocation line.

A common feature of gauge theory, nonlocal elasticity, strain gradient theory 
and theory of Cosserat media of defects is that a characteristic length scale enters the constitutive laws. 
In the gauge theory of defects and in the theory of Cosserat and multipolar media 
this length is a material property which carries 
with it all of the difference between solutions with or without 
moment stresses. 
Its influence might be become important as dimensions of the body or wavelengths 
diminish to the order of the characteristic inner length.
By means of this characteristic length scale
it is possible to define the dislocation core radius in a straightforward 
manner~\cite{Lazar02}.

The question arises: Can we find a gauge theoretical solution for the 
edge dislocation which has the correct far field stress and 
is related to any solution of another framework?
The aim of this paper is to find an answer to this question,
that means to construct a correct gauge theoretical solution of a straight edge dislocation. 
Additionally, the relation to the nonlocal
theory and strain gradient elasticity of an edge dislocation will
be given.
Here we restrict our considerations to the case of linear dislocation theory.

\section{The translational gauge invariant formulation of dislocation theory}
\setcounter{equation}{0}
For the undeformed reference state of the body the indices $i=1,2,3$ are used.
The final coordinates of the deformed state are labelled by the indices
$a=1,2,3$. In presence of defects the deformed state of the material is 
characterized by nonholonomic coordinates.
In linear theory we may use $\delta_a^{\ i}$, $\delta_{ij}$ and $\delta_{ab}$ 
to raise and lower the indices. 

In elasticity theory the state of an elastic body is not changed under a 
rigid translation and rotation (global gauge transformations). 
In elasticity the (compatible) distortion is
given as a proper gradient 
\begin{align}
\label{dist1}
\beta^a_{\ i}=\pd_i u^a.
\end{align}
The distortion tensor is dimensionless.
The basic idea of gauge theory of defects is now that the distortion
has to be invariant not only under the global gauge transformation but
also under if the gauge group is applied locally.
In this way, one obtains the incompatible distortion 
in the framework of translational gauge theory of dislocations as follows
\begin{align}
\label{dist2}
\beta^a_{\ i}=\pd_i u^a+\phi^a_{\ i}.
\end{align}
It is locally translational invariant (gauge invariant) due to
\begin{align}
\label{gauge-trans}
u^a\longrightarrow u^a+\tau^a(x),\qquad 
\phi^a_{\ i}\longrightarrow\phi^a_{\ i}-\pd_i\tau^a(x),
\end{align}
where $\tau^a(x)$ are local translations~\cite{Lazar02}.
Here $u^a$ is the displacement field from the undeformed to
the deformed configuration and $\phi^a_{\ i}$ is the proper
incompatible part of the distortion.
The incompatible distortion (\ref{dist2}) can be understood as the
(minimal) replacement of the compatible distortion (\ref{dist1}) in 
the $T(3)$-gauge theory
\begin{align}
\pd_i u^a\longrightarrow \pd_i u^a+\phi^a_{\ i}.
\end{align}
Then, the (incompatible) strain tensor turns out to be\footnote{We will be using
the notations $A_{(ij)}\equiv\frac{1}{2}(A_{ij}+A_{ji})$ and 
$A_{[ij]}\equiv\frac{1}{2}(A_{ij}-A_{ji})$.}
\begin{align}
E_{ij}=\frac{1}{2}\big(\delta_{ai}\beta^a_{\ j}+\delta_{aj}\beta^a_{\ i} \big)
        \equiv \beta_{(ij)}.
\end{align}
The skew-symmetric part of the distortion defines
the elastic rotation tensor
\begin{align}
\label{rot}
\omega_{ij}=\frac{1}{2}\big(\delta_{ai}\beta^a_{\ j}-\delta_{aj}\beta^a_{\ i} \big)
        \equiv \beta_{[ij]},
\end{align}
and the elastic rotation vector 
\begin{align}
\label{rot1}
\omega_i\equiv-\frac{1}{2}\,\epsilon_i^{\ jk}\delta_{aj}\beta^a_{\ k}.
\end{align}
%%After minimal replacement 
The elastic strain energy of an isotropic media is defined by
\begin{align}
\label{strain-energy}
W=\frac{1}{2}\,\mu\left(\delta^{ik}\delta^{jl}+\delta^{il}\delta^{jk}
+\frac{2\nu}{1-2\nu}\,\delta^{ij}\delta^{kl}\right)E_{ij}E_{kl},
\end{align}
where $\mu$ is the shear modulus and $\nu$ is Poisson's ratio.
It is now invariant under local translations.
Obviously, no elastic rotation enters the relation~(\ref{strain-energy}) 
and, therefore, the elastic rotation has no elastic response quantity.
The force stress can be derived from the strain energy
%%\begin{align}
%%\label{FS-PK}
%%\sigma_a^{\ j}=\frac{\pd W}{\pd \beta^a_{\ j}}.
%%\end{align}
%%%Here $a$ is the force index and the index $j$ charaterizes the normal of the 
%%%surface on which the force acts.
%%%The Cauchy stress tensor is defined by
\begin{align}
\label{FS-sym}
\sigma^{ij}=\frac{\pd W}{\pd E_{ij}}=
2\mu\left(E^{ij}+\frac{\nu}{1-2\nu}\,\delta^{ij} E_k^{\ k}\right).
\end{align}
It is the elastic response quantity to the strain tensor.
Formula~(\ref{FS-sym}) expresses the (incompatible) stress-strain relation 
(generalized Hooke's law). 
%%In linear theory we have not to make a difference
%%between the Piola-Kirchoff and the Cauchy stress tensor. 

It is well known that nontrivial traction boundary problems 
in the variational formulation of the gauge theory of defects
can be formulated by means of a so-called
null Lagrangian~\cite{Edelen88,Edelen96,Edelen89}. 
When the null Lagrangian is added to the Lagrangian of elasticity, it does not change the
Euler-Lagrange equation (force equilibrium) because the 
associated Euler-Lagrange equation, $\pd_j\tl\sigma {}^{\ j}_{a}=0$, is 
identically satisfied.
After minimal replacement, $\pd_i u^a\rightarrow \beta^a_{\ i}$, the null
Lagrangian 
\begin{align}
\label{L-null}
W_{\rm bg}=\pd_j\big(\tl\sigma {}_a^{\ j} u^a\big)
          =\big(\pd_j \tl\sigma {}_{a}^{\ j}\big) u^a
            +\tl\sigma {}^{\ j}_{a} \pd_j u^a
          \longrightarrow\tl\sigma {}^{\ j}_{a} \beta^{a}_{\ j},
\end{align}
give rise to a background stress tensor $\tl\sigma {}^{\ j}_{a}$
which can be considered as the nucleation field in the gauge theory of 
defects~\cite{Edelen96}.

As usual we use the translational field strength (torsion) to define the dislocation density 
tensor~\cite{Kondo52,Bilby55,KS59,Kroener60,Kroener93,Kroener96,Lazar00,Lazar02,Lazar02b}
\begin{align}
T^a_{\ ij}=\pd_i\beta^a_{\ j}-\pd_j\beta^a_{\ i},
\end{align}
which has the dimension of an inverse length.
The dislocation density is a fundamental quantity in plasticity because the
dislocation is the elementary carrier of plasticity.
This field is a physical state quantity
which, at least in principle, can be measured without knowing anything about 
the history of the body. It is possible with the help of high resolution transmission
electron microscopes (HRTEM) to see single dislocations with their 
core configuration in crystals.
Since dislocations change the energy of the body (crystal), they should
appear in the total Lagrangian.
The dislocation Lagrangian, which we use, contains only the translational gauge field strength 
and is given by
\begin{align}
\label{L-core}
{\cal L}_{\rm disl}=
-\frac{1}{4}\, T^a_{\ ij} H_a^{\ ij},
\end{align}
where the moment stress $H_a^{\ ij}$ 
is the response quantity to dislocation density. 
This means, that at all positions where the dislocation density is non zero,
also localized moment stresses are present.
It is convenient to perform an irreducible decomposition of the torsion, 
from which we can construct the most general 
isotropic constitutive law between the dislocation density and the moment 
stress in the following way~\cite{Lazar00,Lazar02} 
\begin{align}
\label{const_iso}
H_{aij}=\sum_{I=1}^{3}a_{I}\,^{(I)}T_{aij}.
\end{align}
Here $a_1$, $a_2$ and $a_3$ are new material coefficients. They have the
dimension of a force.
The irreducible decomposition of the torsion under the rotation group $SO(3)$ 
reads
\begin{align}
T_{aij}=\,^{(1)}T_{aij}+\,^{(2)}T_{aij}+\,^{(3)}T_{aij}.
\end{align}
The tensor, the trace and the axial tensor pieces are defined
by (see also~\cite{MAG,Lazar02})
\begin{alignat}{2}
\label{tentor}
^{(1)}T_{aij}&=T_{aij}-\,^{(2)}T_{aij}-\,^{(3)}T_{aij}
&&\qquad \text{(tentor)},\\
\label{trator}
^{(2)}T_{aij}&:=\frac{1}{2}\left(\delta_{ai}T^l_{\ lj}
                         +\delta_{aj}T^l_{\ il}\right)
&&\qquad\text{(trator)},\\
\label{axitor}
^{(3)}T_{aij}&:=\frac{1}{3}\left(T_{aij}+T_{ija}+T_{jai}\right)
&&\qquad\text{(axitor).}
\end{alignat}
Then, the moment stress tensor reads
\begin{align}
\label{moment1}
H_{aij}=-2\frac{\pd{\cal L}_{\rm disl}}{\pd T^{aij}}
=c_1\, T_{aij} 
+c_2\,\big(T_{ija}+T_{jai}\big)
+c_3\,\big(\delta_{ai}T^l_{\ lj}+\delta_{aj}T^l_{\ il}\big),
\end{align}
with the abbreviations 
\begin{align}
c_1:=\frac{1}{3}\big(2a_1+a_3\big),\quad
c_2:=\frac{1}{3}\big(a_3-a_1\big),\quad
c_3:=\frac{1}{2}\big(a_2-a_1\big).
\end{align}

The Euler-Lagrange equations for the Lagrangian ${\cal L}={\cal L}_{\rm disl}-W+W_{\rm bg}$
turn out to be~\cite{Lazar02}
\begin{align}
\label{feq}
\frac{\delta{\cal L}}{\delta u^a}&\equiv 
\pd_j\sigma_{a}^{\ j}=0& &\text{(force equilibrium)},\\
\label{meq1}
\frac{\delta{\cal L}}{\delta \phi^a_{\ j}}&\equiv 
\pd_i H_{a}^{\ ij} =\widehat\sigma_{a}^{\ j}& &\text{(moment equilibrium)},
\end{align}
with the so-called effective stress tensor
\begin{align}
\widehat\sigma_{a}^{\ j}:=\sigma_{a}^{\ j}-\tl\sigma {}_{a}^{\ j},
\end{align}
which drives the dislocation fields. 
The field equations (\ref{feq}) and (\ref{meq1}) determine the fields
$\beta^a_{\ j}$ and $T^a_{\ ij}$ which represent the geometric (or 
kinematic) degrees of freedom. 
Note that in our approach the dislocation density is not given, a priori, as delta function.
In this field theoretical context, we are able to calculate even the 
dislocation core.
The static responses are the  fields $\sigma_a^{\ j}$
and $H_a^{\ ij}$.

The L.H.S. of~(\ref{meq1}) can be written in terms of the distortion tensor 
\begin{align}
\label{Moment1}
\pd^i H_{aij}
=c_1\big(\Delta\beta_{aj}-\pd^i\pd_j\beta_{ai}\big)
&+c_2\big(\pd^i\pd_j\beta_{ia}-\pd_a\pd^i\beta_{ij}
        +\pd_a\pd^i\beta_{ji}-\Delta\beta_{ja}\big)\\
&+c_3\big(\pd_a\pd^i\beta_{ij}-\pd_a\pd_j\beta^i_{\ i}
        +\delta_{aj}\Delta\beta^i_{\ i}-\delta_{aj}\pd^i\pd^k\beta_{ik}\big).
\nonumber
\end{align}
In the general case the gauge theory of dislocations for isotropic materials 
contains three new material coefficients 
which can be used to define three internal characteristic lengths.
In order to obtain only one coefficient from the three ones 
we need a suitable choice for the moment stress tensor $H_{aij}$ 
and the coefficients $a_1$, $a_2$ and $a_3$, respectively.
By means of such a choice we want to find a gauge theoretical solution 
of symmetric force stresses for the edge dislocation.
If we use the choice\footnote{Note that Malyshev~\cite{Malyshev00}
used the so-called Einstein choice $a_2=-a_1$ and $a_3=-\frac{a_1}{2}$
for an (modified) edge dislocation. 
By applying the stress function method he found a gauge theoretical solution
for the edge dislocation in terms of the Bessel and Neumann functions
and the modified Bessel functions. 
His solution of an (modified) edge dislocation has no singularity at $r=0$ but 
differs asymptotically from the classical edge dislocation. 
In detail the symmetric stress field, the closure failure and the dislocation density 
show a decaying behaviour with oscillations. 
Additionally, his solution violates in the dislocation core region 
the plane-strain condition $E_{zz}=0$.}
\begin{align}
\label{choice-L}
a_2=\frac{1+\nu}{1-\nu}\, a_1,\quad a_3=-\frac{a_1}{2},
\quad \Longrightarrow\quad
c_1=\frac{a_1}{2},\quad c_2=-\frac{a_1}{2},\quad c_3=\frac{\nu}{1-\nu}\, a_1,
\end{align}
the localized moment stress tensor reads
\begin{align}
\label{moment2}
H_{aij}=\frac{a_1}{2}\left( T_{aij}-T_{ija}-T_{jai}
+\frac{2\nu}{1-\nu}\,\big(\delta_{ai}T^l_{\ lj}+\delta_{aj}T^l_{\ il}
\big)\right).
\end{align}
With the relations 
$H_{ak}=\frac{1}{2}\epsilon_{ijk} H_a^{\ ij}$ and
$T^a_{\ ij}=\epsilon_{ijk} \alpha^{ak}$, 
and by using
the conventional (microscopic) dislocation density 
tensor $\alpha^{ai}=\epsilon^{ijk}\pd_j\beta^a_{\ k}$ 
and the Nye~\cite{Nye} tensor\footnote{Here the tensor $\kappa_{aj}$ is a 
microscopic quantity.} $\kappa_{aj}$ 
\begin{align}
\kappa_{aj}=\alpha_{ja}-\frac{1}{2}\, \delta_{aj}\alpha^k_{\ k},\qquad
\alpha_{aj}=\kappa_{ja}-\delta_{aj}\kappa^k_{\ k},
\end{align}
we obtain from (\ref{moment1}) the relation
\begin{align}
\label{moment-2}
H_{aj}= a_1\left(\alpha_{aj}-\frac{1}{2}\,\delta_{aj}\alpha^k_{\ k}
+\frac{2\nu}{1-\nu}\,\alpha_{[aj]}\right)
        = a_1\left(\kappa_{ja}+\frac{2\nu}{1-\nu}\,\kappa_{[ja]}\right).
\end{align}
Alternatively, the dislocation Lagrangian~(\ref{L-core}) can be written 
in the form
\begin{align}
{\cal L}_{\rm disl}=-\frac{1}{2}\, \alpha^{ai} H_{ai}.
\end{align}
By using~(\ref{moment-2}) the L.H.S. of~(\ref{meq1}) reads
\begin{align}
\label{fel1}
\pd^i H_{aij}
=a_1&\bigg\{\Delta\beta_{(aj)}-\pd^i\pd_j\beta_{(ai)}
+\frac{1}{1-\nu}\, \pd_a\pd^i\beta_{[ij]}\\
&-\frac{\nu}{1-\nu}\Big(\big(\pd_a\pd_j-\delta_{aj}\Delta\big)\beta^i_{\ i}
+\delta_{aj}\pd^i\pd^k\beta_{ik}-\pd_a\pd^i\beta_{(ij)}\Big)\bigg\}.
\nonumber
\end{align}
Obviously, the strain and the rotation tensor appear in~(\ref{fel1})
because the distortion is not symmetric. 
It is reminded that in the so-called Hilbert-Einstein choice only the strain 
tensor enters the moment equilibrium (see~\cite{Lazar02,Lazar02b}).

We now define the rotation gradient\footnote{DeWit~\cite{deWit70,deWit73} 
called such a rotation gradient elastic bend-twist.}
\begin{align}
\label{curv1}
k^a_{\ ij}=\pd^a\beta_{[ij]}.
\end{align}
It has the dimension of reciprocal of length.
The rotation gradient describes a ``lattice-curvature''. 
It contains two pieces. To see this we rewrite~(\ref{curv1}) 
\begin{align}
\label{curv2}
k_{ijk}=\big(\pd_k E_{ij}-\pd_j E_{ik}\big)
	+\frac{1}{2}\big(T_{ijk}-T_{jki}-T_{kij}\big).
\end{align}
The first one is caused by local (incompatible) elastic strain 
and the other one by dislocations (contortion)~\cite{Kroener60}. 
The second piece may be identified with the plastic ``lattice-curvature''
which cannot be realized without dislocations.
We add the following gauge invariant Lagrangian 
which looks like a rotation gradient Lagrangian of a Cosserat 
medium (see, e.g.,~\cite{Mindlin65})
\begin{align}
\label{L-grad}
{\cal L}_{\rm grad}=-\frac{1}{2}\, k^a_{\ ij}\mu_a^{\ ij}.
\end{align}
The response quantity to the rotation gradient is 
a so-called ``Cosserat'' couple stress for an isotropic 
material\footnote{The relations between the constants in~(\ref{moment-mu}) and
those in~\cite{Mindlin65} are: $d_1=2\alpha_2$, $d_2=\alpha_3$, $d_3=-\alpha_1$.}
(see, e.g.,~\cite{Mindlin65})
\begin{align}
\label{moment-mu}
\mu_{a[ij]}=-\frac{\pd{\cal L}_{\rm grad}}{\pd k^{a[ij]}}
           =d_1 k_{a[ij]} +d_2\big(k_{i[ja]}+k_{j[ai]}\big)
	    +d_3\big(\delta_{ai} k^l_{\ [lj]}+\delta_{aj} k^l_{\ [il]}\big).	
\end{align}
This constitutive relation uses the three rotation gradients, 
which are invariant under the $SO(3)$, and 
is analogous to the invariant constitutive relation for
the torsion~(\ref{moment1}). 
We find
\begin{align}
\pd^i\mu_{i[aj]}=d_1 \Delta \beta_{[aj]}
+(d_2-d_3)\big(\pd^i\pd_j\beta_{[ia]}-\pd^i\pd_a\beta_{[ij]} \big).
\end{align}

Another gauge invariant Lagrangian, which we add, is given in terms 
of the torsion and the rotation gradient 
\begin{align}
\label{L-int}
{\cal L}_{\rm grad-disl}=-\frac{1}{2}\,d_0 k^a_{\ ij} T_a^{\ ij}.
\end{align}
Equation~(\ref{L-int}) may be considered as a kind of an interaction Lagrangian 
between dislocations and the ``lattice-curvature''.
We define the response quantity
\begin{align}
\label{moment-lam}
\lambda_{aij}=-\frac{\pd{\cal L}_{\rm grad-disl}}{\pd (\pd^i\beta^{aj})}
           =d_0\Big( k_{aij} +\frac{1}{2}\,T_{iaj}\Big),
\end{align}
with
\begin{align}
\pd^i\lambda_{aij}=d_0\, \pd^i\Big(\pd_a\beta_{[ij]}+\frac{1}{2}\big(\pd_a\beta_{ij}-\pd_j\beta_{ia}\big)\Big).
\end{align}
The coefficients $d_0$, $d_1$, $d_2$ and $d_3$ have the dimensions of forces.

If we use the Lagrangian ${\cal L}={\cal L}_{\rm disl}+{\cal L}_{\rm grad}
+{\cal L}_{\rm grad-disl}-W+W_{\rm bg}$,
the moment equilibrium reads
\begin{align}
\label{meq2}
\pd^i\left(H_{aij}+\lambda_{aij}+\mu_{i[aj]}\right)=\widehat\sigma_{aj},
\end{align}
which can be decomposed into its symmetric and antisymmetric part.
For the symmetric part of the moment equilibrium we obtain the equation 
\begin{align}
\label{moment2b}
\pd^i\left(H_{(aj)i}+\lambda_{(aj)i}\right)+\widehat\sigma_{(aj)}=0.
\end{align}
By using the choice of the coefficient
\begin{align}
\label{choice-d0}
d_0=-\frac{a_1}{(1-\nu)},
\end{align}
formula~(\ref{moment2b}) can be rewritten in terms of the strain tensor,
only, according to
\begin{align}
\label{moment3}
a_1\bigg\{\Delta E_{ij}-\pd^k\pd_{(i} E_{j)k}
-\frac{\nu}{1-\nu}\Big(\big(\pd_i\pd_j-\delta_{ij}\Delta\big)E^k_{\ k}
+\delta_{ij}\pd^k\pd^l E_{kl}-\pd^k\pd_{(i} E_{j)k}\Big)\bigg\}
=\widehat\sigma_{ij}.
\end{align}
This equation can be interpreted as the (static) field equation 
in the elastoplastic theory of 
dislocations.
It is derived from the combination of several Lagrangians
 -- Eqs.~(\ref{L-core}), (\ref{L-grad}), (\ref{L-int}) -- 
and the consideration of the null Lagrangian~(\ref{L-null}) and 
the strain energy~(\ref{strain-energy}) in combination with 
the special choices of material constants~(\ref{choice-L}) and 
(\ref{choice-d0}).
The antisymmetric part of the moment 
equilibrium\footnote{Indices which are exempt from antisymmetrisation 
are enclosed by vertical bars.}
\begin{align}
\label{moment4}
\pd^i\left(H_{[aj]i}-\lambda_{[a|i|j]}-\mu_{i[aj]}\right)=0,
\end{align}
can be satisfied by the help of the choice of the coefficients
\begin{align}
\label{choice-d1}
d_1=0,\qquad d_2-d_3=\frac{d_0}{2}.
\end{align}
The physical interpretation of Eq.~(\ref{moment4}) is the following.
Because we have required that the force stress should be symmetric, $\sigma_{[ij]}=0$, 
we have had to introduce the couple stresses $\lambda_{[a|i|j]}$ and $\mu_{i[aj]}$ in order 
to fulfill the antisymmetric moment equilibrium~(\ref{moment4}). 
Consequently, (\ref{moment4}) expresses the moment equilibrium between
the antisymmetric moment stresses $H_{[aj]i}$, $\lambda_{[a|i|j]}$ and 
$\mu_{i[aj]}$. 
The fact that the moment stress is the specific 
response to the presence of dislocations and that the field 
equation~(\ref{meq2}) expresses the moment equilibrium is a hint 
that the gauge theory of dislocations may be connected with the theory
of a constrained, anholonomic Cosserat media with symmetric force stress.
The dislocations bring the anholonomity into the picture.
A similar remark in this direction was given by Kluge~\cite{Kluge69}.
It should be also possible to extend this theory 
to a more general dislocation theory in a
general Cosserat continuum with antisymmetric force stress and 
10 material constants 
which is not the aim of the present paper.
However, for antisymmetric force stress and vanishing dislocation density
the moment equilibrium equation recovers the form of the Cosserat equilibrium
equation under zero external moments.

Finally, we may define the hyperstress 
\begin{align}
\label{hyper-stress}
M_a^{\ ij}\equiv-\frac{\pd{\cal L}}{\pd(\pd_i\beta^a_{\ j})}=
	   H_a^{\ ij}+\lambda_a^{\ ij}+\mu_{\, a}^{i\ j}.
\end{align}
This hyperstress tensor is close to Maugin's so-called ``Piola-Kirchhoff'' 
hyperstress~\cite{Maugin}. Such a hyperstress tensor was also used by 
Toupin for the theory of nonsimple elastic materials~\cite{Toupin64}.
Thus, the dislocation theory given in this paper 
can be considered as an anholonomic higher-order gradient theory in nonsimple 
materials. 
If $a_1=0$, there are no effects of hyperstresses and couple stresses.

By the help of the inverse of Hooke's law
\begin{align}
\label{hooke-inv}
E_{ij}=\frac{1}{2\mu}\left(\sigma_{ij}-\frac{\nu}{1+\nu}\,\delta_{ij}\sigma^k_{\ k}\right)
\end{align}
and using the equilibrium condition $\pd^j\sigma_{ij}=0$, 
we obtain from Eq.~(\ref{moment3}) the inhomogeneous Helmholtz equation
as field equation for every component of the stress tensor
\begin{align}
\label{stress-fe}
\Big(1-\kappa^{-2}\Delta\big)\sigma_{ij}=\tl\sigma {}_{ij},\qquad \kappa^2=\frac{2\mu}{a_1}.
\end{align}
It is important to note that (\ref{stress-fe}) agrees with the field equation
for the stress field in nonlocal elasticity~\cite{Eringen83,Eringen85,Eringen87}
and in gradient elasticity~\cite{GA99,Gutkin00}.
The factor $\kappa^{-1}$ has the physical dimension of a length and 
therefore it defines an internal characteristic length 
(dislocation length scale).
By using~(\ref{hooke-inv}) and (\ref{stress-fe}) we obtain an inhomogeneous
Helmholtz equation for the strain fields 
\begin{align}
\label{strain-fe}
\Big(1-\kappa^{-2}\Delta\big)E_{ij}=\tl E {}_{ij},
\end{align}
where $\tl E {}_{ij}$ is the background strain tensor.
Equation~(\ref{strain-fe}) is similar to the equation for strain of the 
gradient theory used by Gutkin and Aifantis~\cite{GA96,GA97,GA99,Gutkin00},
if we identify $\kappa^{-2}$ with their corresponding gradient coefficient
(see, e.g., equation~(4) in~\cite{GA99}).
Note that only for compatible strain, $\phi_{ij}=0$, one can deduce an 
inhomogeneous Helmholtz equation for the displacement from (\ref{strain-fe}) 
which is analogous to the equation for the displacement in gradient elasticity
(see equation~(3) in~\cite{GA99}).

The conditions on a gauge theoretical solution of 
dislocations in this framework are:
(i)~the stress field should have no singularity at $r=0$,
(ii)~the far field stress ought to be the classical stress field. 
We use the condition (ii) because the classical stress fields are 
well-established and in good agreement with physical observations,
e.g. the investigation of stresses around dislocations by means of the
photoelasticity method (see, e.g.,~\cite{BA56,Bullough58,ND70}). 

\section{The gauge theoretical solution of the straight edge dislocation}
\setcounter{equation}{0}
\subsection{The stress, strain and displacement field in an infinitely 
extended body}
Let us now derive the correct dislocation fields within the translational
gauge theory.
We take Cartesian coordinates, so that the $z$-axis is along the dislocation line
and the $x$-axis is along the Burgers vector: $b^x=b$, $b^y=b^z=0$. 
The extra half plane lies in the plane $x=0$. 
In order to satisfy the equilibrium condition, we use the second order
stress function $f$
%%%tensor $\chi_{ij}$ 
%%%\begin{align}
%%%\sigma_{ij}=\left(\text{inc}\, \boldsymbol{\chi}\right)_{ij}
%%%\end{align}
and specialize to the plane problem of an edge dislocation by
setting $\pd_z\equiv 0$. 
%% and using
%%\begin{align}
%%f\equiv-\chi_{zz},\qquad
%%p\equiv-\pd^2_{yy}\chi_{xx}-\pd^2_{xx}\chi_{yy}+2\pd^2_{xy}\chi_{xy}
%%\end{align}
We are using the so-called stress function ansatz~\cite{HL}
\begin{align}
\label{stress-ansatz}
\sigma_{ij}=
\left(\begin{array}{ccc}
\pd^2_{yy}f & -\pd^2_{xy}f & 0\\
-\pd^2_{xy}f & \pd^2_{xx}f & 0\\
0& 0& p
\end{array}\right),
\end{align}
with 
\begin{align}
\label{p}
p=\nu\Delta f,
\end{align}
where $\Delta$ denotes the two-dimensional Laplacian $\pd^2_{xx}+\pd^2_{yy}$.
The relation~(\ref{p}) comes from the requirement: $E_{zz}=0$ (plane strain).
Additionally, the strain is given in terms of the stress function as
\begin{align}
\label{strain-ansatz}
E_{ij}=\frac{1}{2\mu}
\left(\begin{array}{ccc}
\pd^2_{yy}f-\nu\Delta f & -\pd^2_{xy}f & 0\\
-\pd^2_{xy}f & \pd^2_{xx}f-\nu\Delta f & 0\\
0& 0& 0
\end{array}\right).
\end{align}
%%%%%It is important to note that (\ref{strain-ansatz}) fulfils the antisymmetric
%%%%%moment equilibrium~(\ref{moment5}).
We use the classical stress field of a straight edge dislocation 
in terms of the Airy stress function
as the background stress
\begin{align}
\label{stress-ansatz2}
\tl\sigma {}_{ij}=
\left(\begin{array}{ccc}
\pd^2_{yy}\chi & -\pd^2_{xy}\chi & 0\\
-\pd^2_{xy}\chi & \pd^2_{xx}\chi & 0\\
0& 0& \nu\Delta\chi
\end{array}\right).
\end{align}
The Airy stress function~\cite{Timpe05,Brown41,Koehler41,LL49,Kroener58,Kroener81}
\begin{align}
\label{Airy1}
\chi=-A\, y\ln r,\qquad A=\frac{\mu b}{2\pi(1-\nu)},
\end{align}
fulfills the following inhomogeneous bipotential (or biharmonic) equation 
\begin{align}
\Delta\Delta\,\chi=-4\pi A\,\pd_y\delta(r),
\end{align}
where $\delta(r)$ is the two-dimensional Dirac delta function and
$r^2=x^2+y^2$.
Substituting (\ref{stress-ansatz}) and (\ref{stress-ansatz2})
into (\ref{stress-fe}) we get 
\begin{align}
\label{f_fe}
\Big(\Delta-\kappa^2\Big)f=\kappa^2 A\, y\ln r .
\end{align}
We make the ansatz
\begin{align}
f=-A\,y\ln r +f_{(1)}
\end{align}
and obtain
\begin{align}
\label{f_fe1}
\Big(\Delta-\kappa^2\Big)f_{(1)}=2A\,\pd_y \ln r.
\end{align}
Now we set
\begin{align}
f_{(1)}=2A\,\pd_y f_{(2)}
\end{align}
with
\begin{align}
\label{f_fe2}
\Big(\Delta-\kappa^2\Big)f_{(2)}=\ln r
\end{align}
and we get the solution
\begin{align}
f_{(2)}=-\frac{1}{\kappa^2}\Big(\ln r+K_0(\kappa r)\Big).
\end{align}
Finally, we obtain the solution of (\ref{f_fe})
\begin{align}
f=-A\left(y\ln r+\frac{2}{\kappa^2}\,\pd_y\Big(\ln r+K_0(\kappa r)\Big)\right).
\end{align}
Then the solution of the modified stress function of an edge dislocation
with Burgers vector $b\| x$
reads\footnote{For an edge dislocation with the Burgers vector $b\| y$ the 
modified stress function is given by:
\begin{align}
f=\frac{\mu b}{2\pi(1-\nu)}\, x \left\{\ln r 
+\frac{2}{\kappa^2 r^2}\Big(1-\kappa r K_1(\kappa r)\Big)\right\}.
%%%%%% +\text{const}.
\nonumber
\end{align}}
\begin{align}
\label{f_edge1}
f=-\frac{\mu b}{2\pi(1-\nu)}\, y \left\{\ln r 
+\frac{2}{\kappa^2 r^2}\Big(1-\kappa r K_1(\kappa r)\Big)\right\}, 
%%%%%%%%%+\text{const}
\end{align}
where the first piece is the Airy stress function and $K_n$ is the modified Bessel
function of the second kind and $n=0,1,\ldots$ denotes the order of this function.

By means of Eqs.~(\ref{stress-ansatz}) and (\ref{f_edge1}),
the modified stress of a straight edge dislocation is given by
\begin{align}
\label{T_xx}
&\sigma_{xx}=-\frac{\mu b}{2\pi(1-\nu)}\, 
\frac{y}{r^4}\bigg\{\big(y^2+3x^2\big)+\frac{4}{\kappa^2r^2}\big(y^2-3x^2\big)\\
&\hspace{5cm}
-2 y^2\kappa r K_1(\kappa r)-2\big(y^2-3x^2\big) K_2(\kappa r)\bigg\}\nonumber\\
\label{T_yy}
&\sigma_{yy}=-\frac{\mu b}{2\pi(1-\nu)}\, 
\frac{y}{r^4}\bigg\{\big(y^2-x^2\big)-\frac{4}{\kappa^2r^2}\big(y^2-3x^2\big)\\
&\hspace{5cm}
-2 x^2\kappa r K_1(\kappa r)+2\big(y^2-3x^2\big) K_2(\kappa r)\bigg\},\nonumber\\
\label{T_xy}
&\sigma_{xy}=\frac{\mu b}{2\pi(1-\nu)}\, 
\frac{x}{r^4}\bigg\{\big(x^2-y^2\big)-\frac{4}{\kappa^2r^2}\big(x^2-3y^2\big)\\
&\hspace{5cm}
-2 y^2\kappa r K_1(\kappa r)+2\big(x^2-3y^2\big) K_2(\kappa r)\bigg\},\nonumber\\
\label{T_zz}
&\sigma_{zz}=-\frac{\mu b\nu }{\pi(1-\nu)}\, 
\frac{y}{r^2}\Big\{1-\kappa r K_1(\kappa r)\Big\}.
\end{align}
The stress $\sigma_{zz}$ satisfies the
condition $\sigma_{zz}=\nu(\sigma_{xx}+\sigma_{yy})$. 
The trace of the stress tensor $\sigma^k_{\ k}$ produced by the edge dislocation in
an isotropic medium is
\begin{align}
\label{hyd_p}
\sigma^k_{\ k}=-\frac{\mu b(1+\nu)}{\pi(1-\nu)}\, 
\frac{y}{r^2}\Big\{1-\kappa  r K_1(\kappa r)\Big\}.
\end{align}
%%%%The physical meaning of Eq.~(\ref{hyd_p}) is very simple.
%%%%Over the $z$--$x$-plane the medium is compressed and 
%%%%below the $z$--$x$-plane the medium is dilated.
\begin{figure}[t]\unitlength1cm
\centerline{
\epsfig{figure=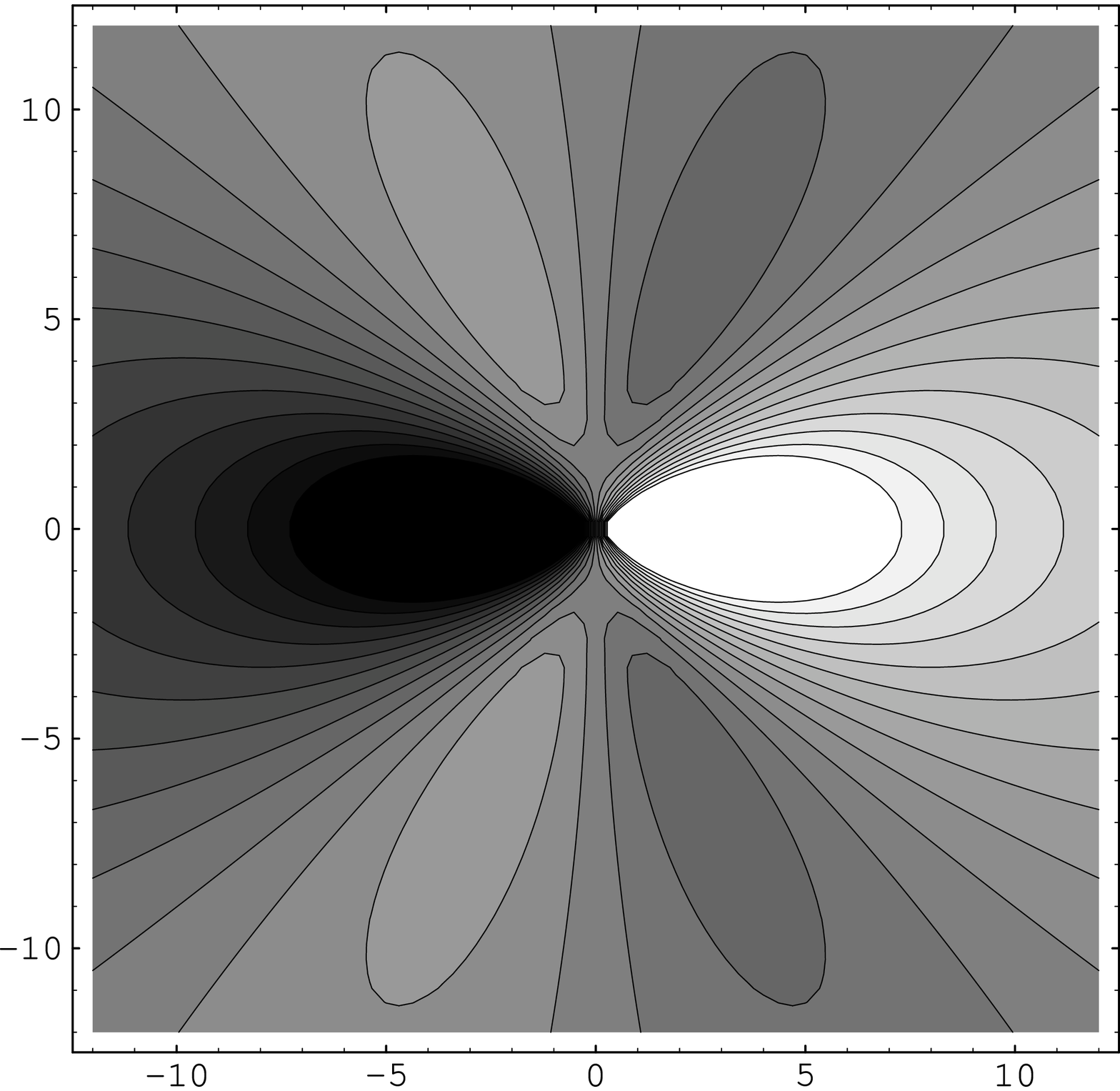,width=6.0cm}
\put(-6.3,3.0){$\kappa y$}
\put(-6.2,-0.3){$\text{(a)}$}
\hspace*{0.2cm}
\put(0,-0.3){$\text{(b)}$}
\epsfig{figure=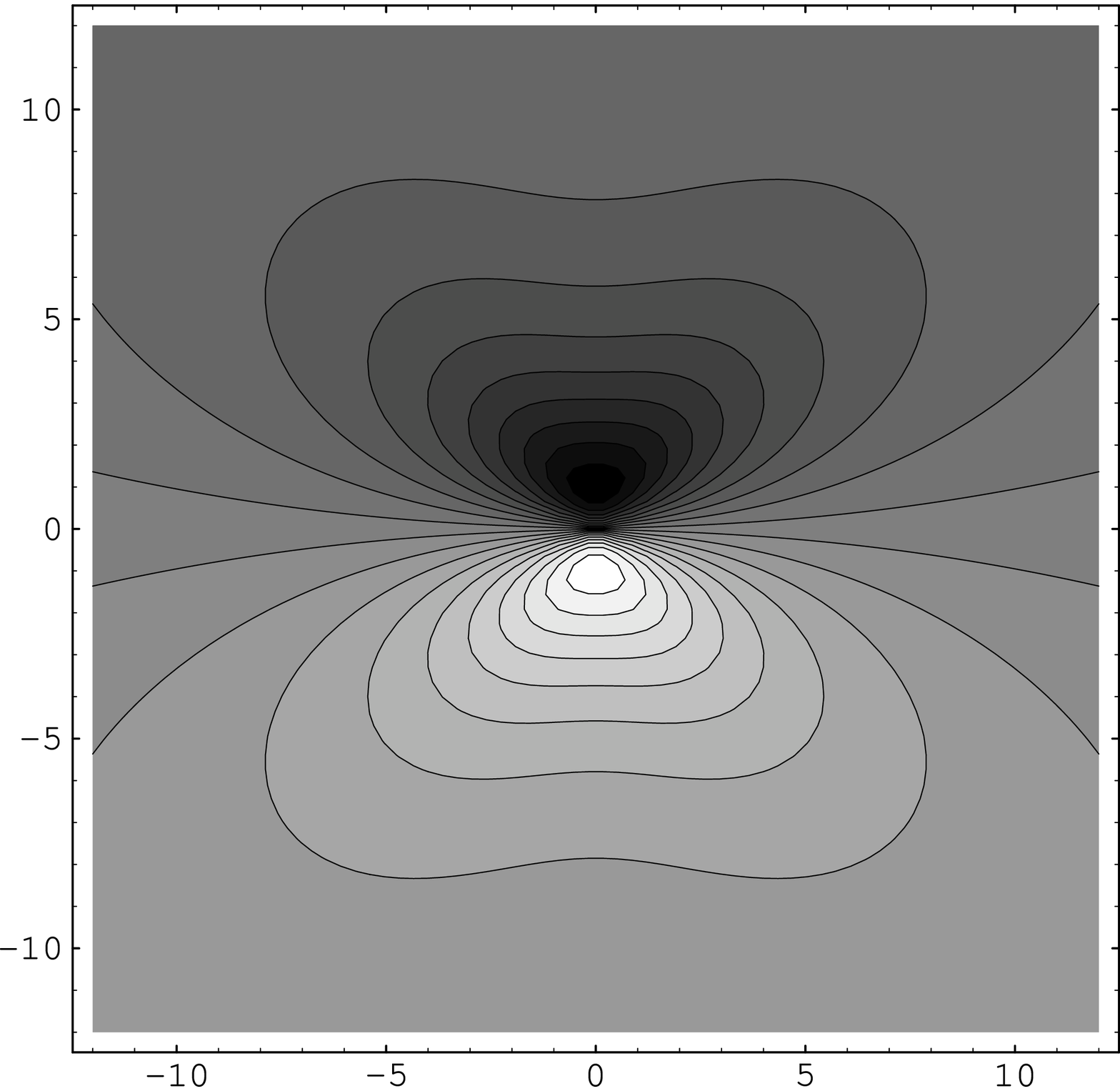,width=6.0cm}
}
\vspace*{0.2cm}
\centerline{
\epsfig{figure=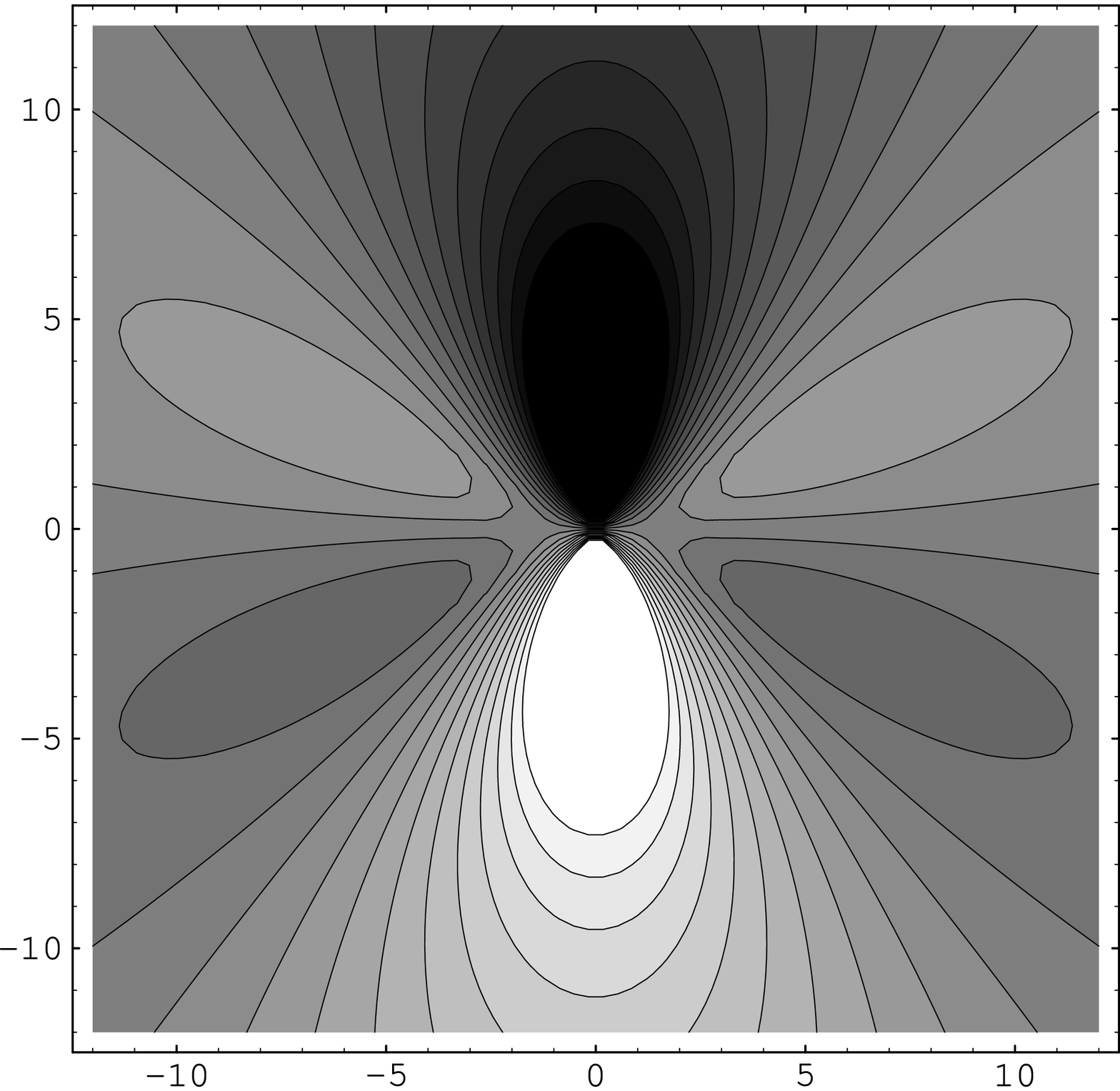,width=6.0cm}
\put(-3.0,-0.3){$\kappa x$}
\put(-6.3,3.0){$\kappa y$}
\put(-6.0,-0.3){$\text{(c)}$}
\hspace*{0.2cm}
\put(3.0,-0.3){$\kappa x$}
\put(0,-0.3){$\text{(d)}$}
\epsfig{figure=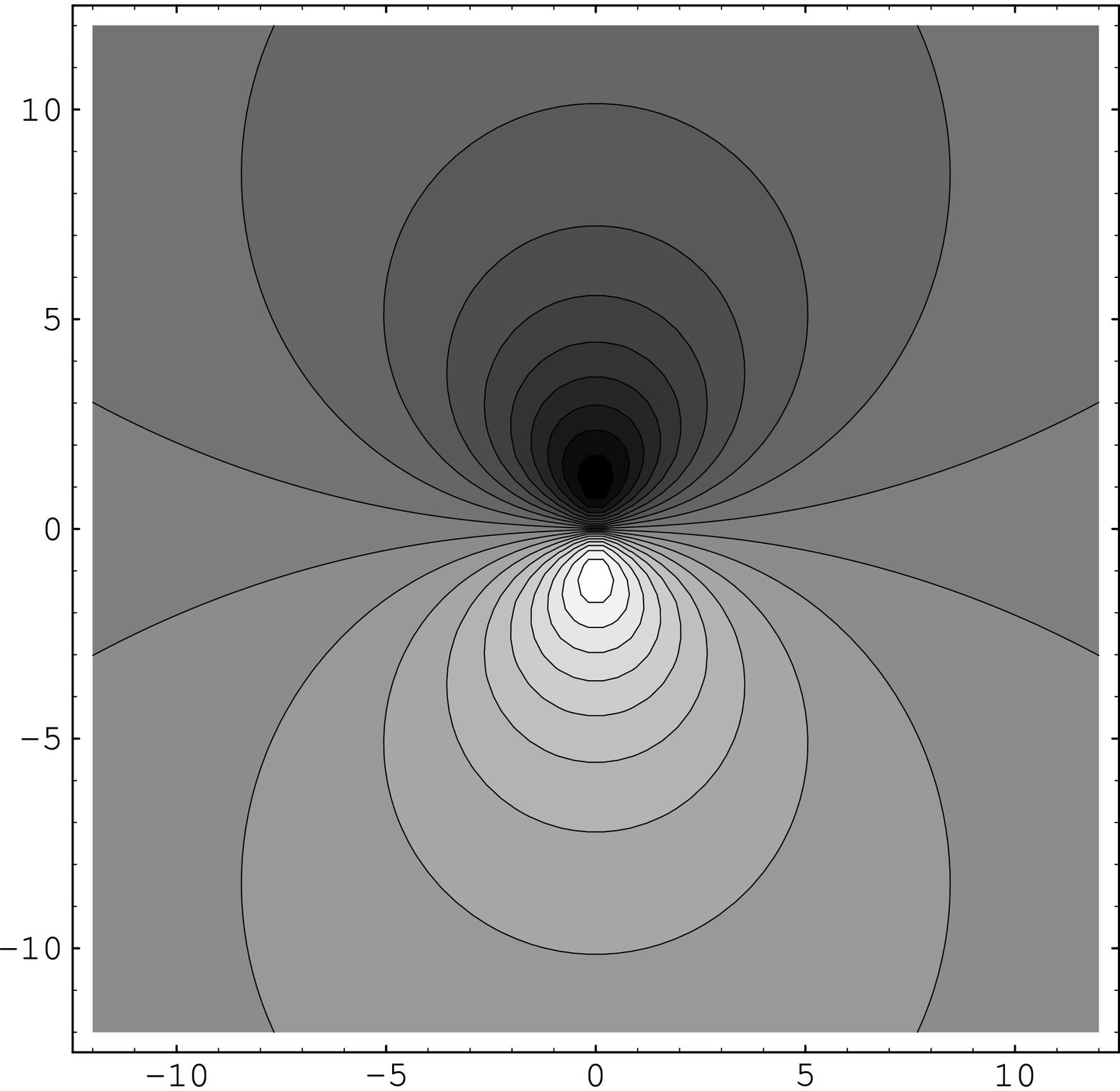,width=6.0cm}
}
\caption{Stress contours of an edge dislocation near the dislocation line:
(a) $\sigma_{xy}$,
(b) $\sigma_{xx}$,
(c) $\sigma_{yy}$,
(d) $\sigma_{zz}$
.}
\label{fig:stress1}
\end{figure}
\begin{figure}[tp]\unitlength1cm
\vspace*{-1.0cm}
\centerline{
(a)
\begin{picture}(8,6)
\put(0.0,0.2){\epsfig{file=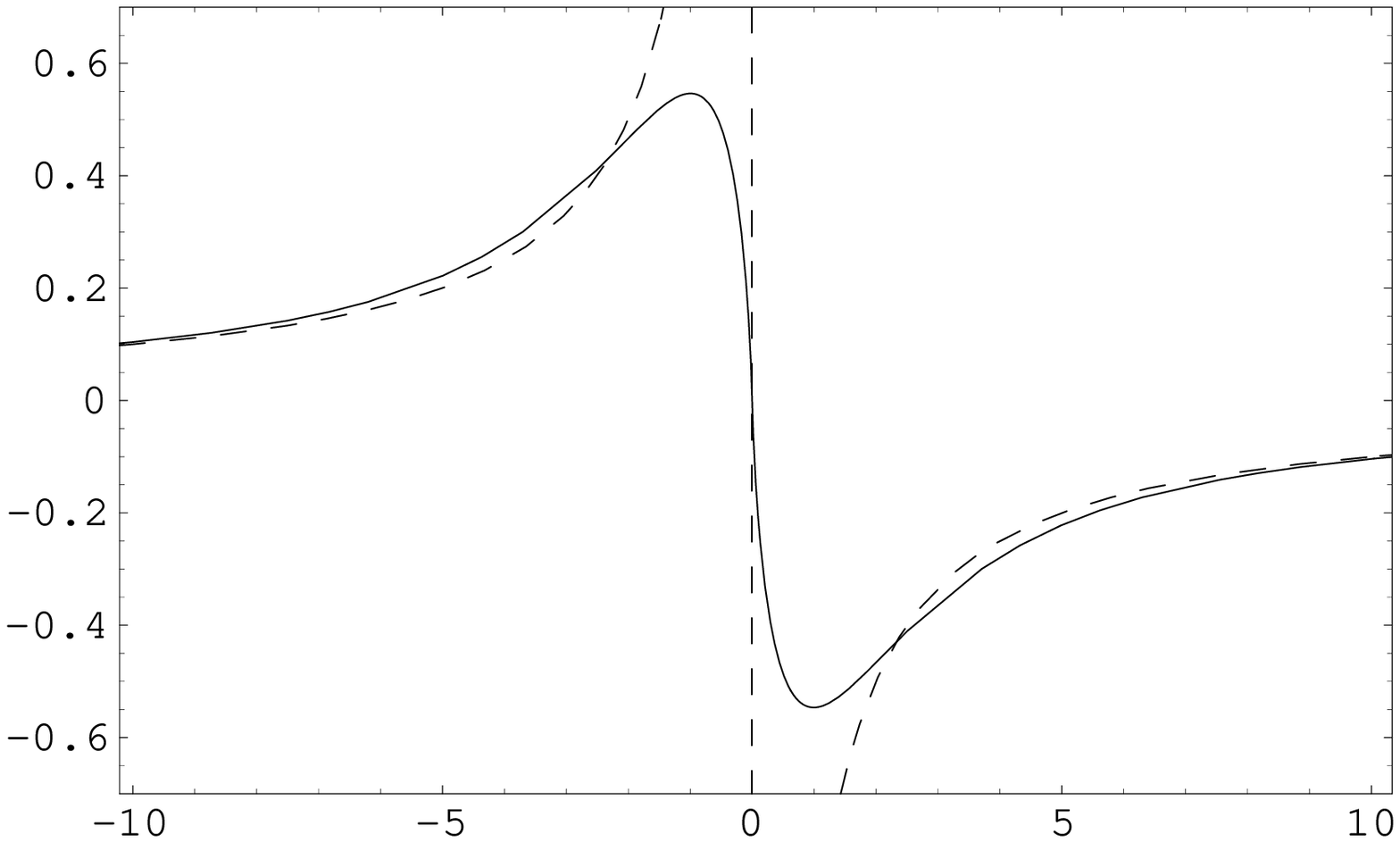,width=8cm}}
\put(4.0,0.0){$\kappa y$}
\put(-1.5,4.0){$\sigma_{xx}(0,y)$}
\end{picture}
}
%%\caption{Displacement field $u^x/b$.}
%%\label{fig:u_x}
%%\end{figure}
%%\begin{figure}[t]\unitlength1cm
\vspace*{-1.0cm}
\centerline{
(b)
\begin{picture}(8,6)
\put(0.0,0.2){\epsfig{file=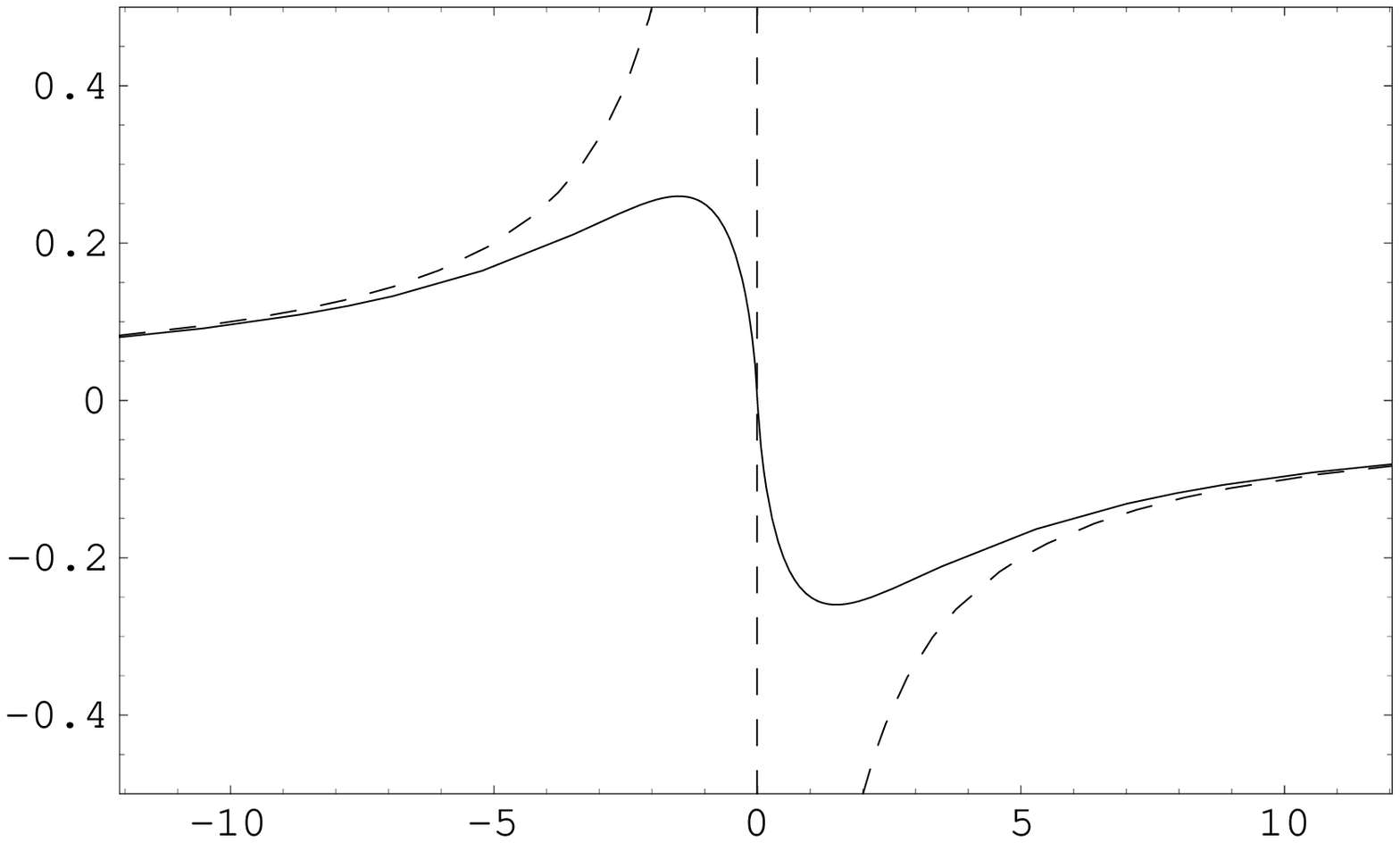,width=8cm}}
\put(4.0,0.0){$\kappa y$}
\put(-1.5,4.0){$\sigma_{yy}(0,y)$}
\end{picture}
}
\vspace*{-1.0cm}
\centerline{
(c)
\begin{picture}(8,6)
\put(0.0,0.2){\epsfig{file=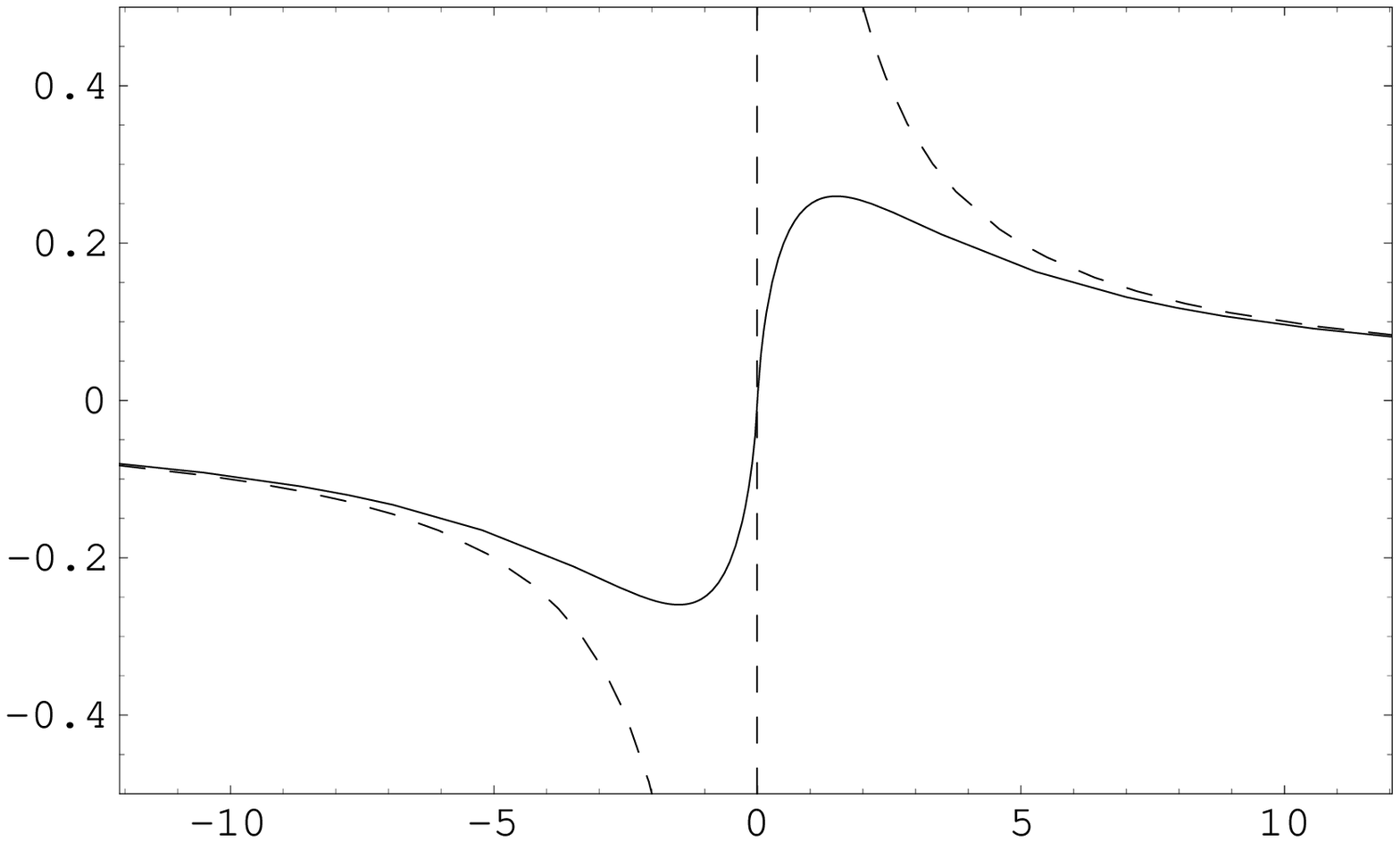,width=8cm}}
\put(4.0,0.0){$\kappa x$}
\put(-1.5,4.0){$\sigma_{xy}(x,0)$}
\end{picture}
}
\vspace*{-1.0cm}
\centerline{
(d)
\begin{picture}(8,6)
\put(0.0,0.2){\epsfig{file=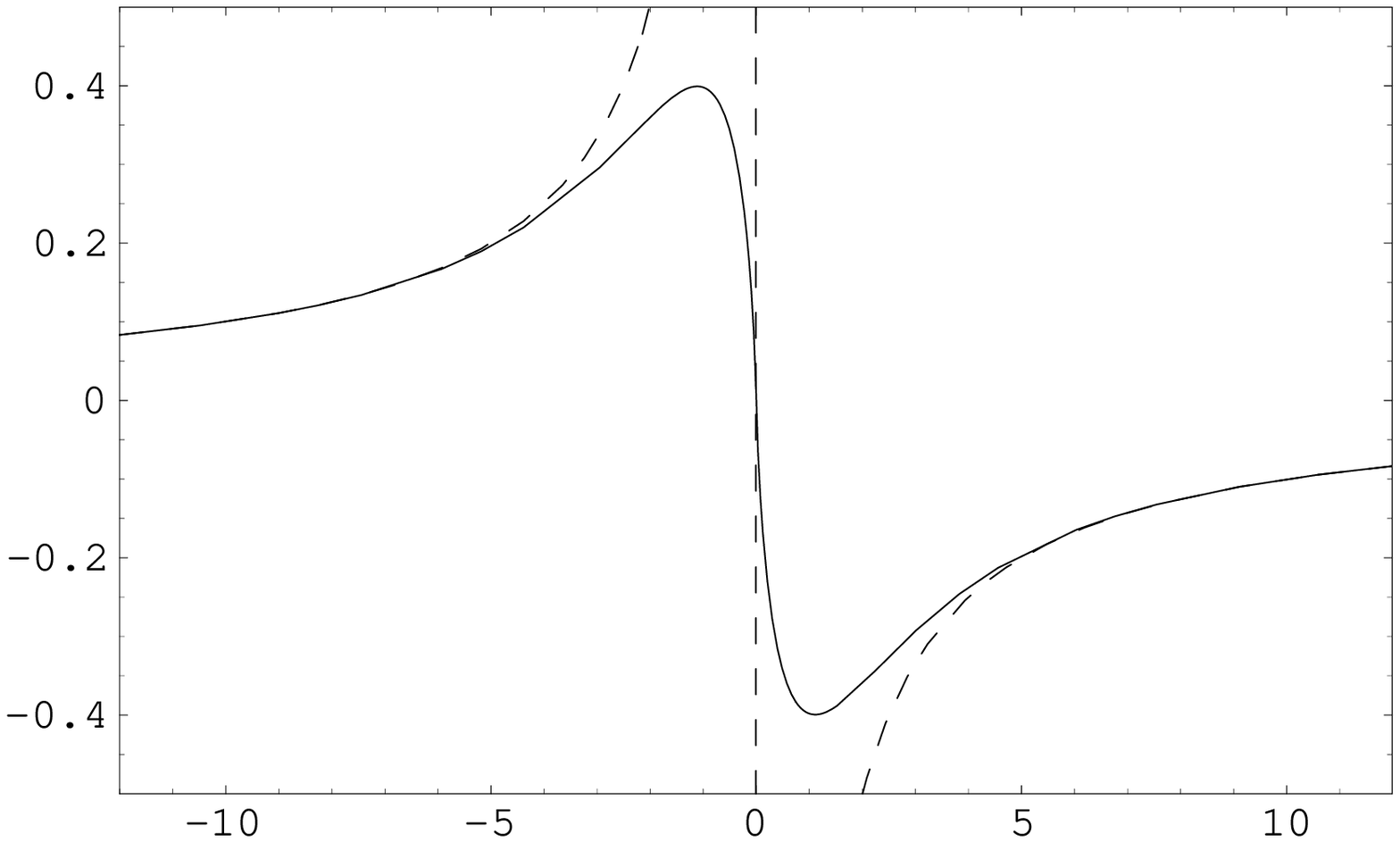,width=8cm}}
\put(4.0,0.0){$\kappa y$}
\put(-1.5,4.0){$\sigma_{zz}(0,y)$}
\end{picture}
}
\caption{The stress components near the dislocation line: 
(a) $\sigma_{xx}(0,y)$, (b) $\sigma_{yy}(0,y)$, 
(c) $\sigma_{xy}(x,0)$ are given in units of $\mu b\kappa/[2\pi(1-\nu)]$ 
and 
(d) $\sigma_{zz}(0,y)$ is given in units of $\mu b\nu \kappa/[\pi(1-\nu)]$. 
The dashed curves represent the classical stress components.}
\label{fig:stress2}
\end{figure}
The spatial distribution of stresses of an edge dislocation 
near the dislocation line are presented in Fig.~\ref{fig:stress1}.

Let us now discuss some details of the core modification of the stress fields.
The stress fields have no artificial singularities at the core  
and the maximum stress occurs at a short distance away from 
the dislocation line (see~ Fig.~\ref{fig:stress2}). 
In fact, when $r\rightarrow 0$, we have
\begin{align}
%%%%K_0(\kappa r)\approx-\gamma-\ln \frac{\kappa r}{2},\quad
K_1(\kappa r)\rightarrow \frac{1}{\kappa r},\qquad
K_2(\kappa r)\rightarrow -\frac{1}{2}+\frac{2}{(\kappa r)^2},
\end{align}
and thus $\sigma_{ij}\rightarrow 0$.
It can be seen that the stresses have the
following extreme values:
$|\sigma_{xx}(0,y)|\simeq 0.546\kappa \frac{\mu b}{2\pi(1-\nu)}$ at 
$|y|\simeq 0.996 \kappa^{-1}$,
$|\sigma_{yy}(0,y)|\simeq 0.260 \kappa\frac{\mu b}{2\pi(1-\nu)}$ at 
$|y|\simeq 1.494 \kappa^{-1}$,
$|\sigma_{xy}(x,0)|\simeq 0.260 \kappa\frac{\mu b}{2\pi(1-\nu)}$ at 
$|x|\simeq 1.494 \kappa^{-1}$,
and
$|\sigma_{zz}(0,y)|\simeq 0.399\kappa \frac{\mu b\nu}{\pi(1-\nu)}$ at 
$|y|\simeq 1.114 \kappa^{-1}$.
The stresses $\sigma_{xx}$,  $\sigma_{yy}$ and  $\sigma_{xy}$ are modified 
near the dislocation core ($0\le r\le 12\kappa^{-1}$). 
Note that $|\sigma_{xy}(x,0)|\simeq 0.260 \kappa\frac{\mu b}{2\pi(1-\nu)}$
can be identified as the theoretical shear strength.
The stress
$\sigma_{zz}$ and the trace $\sigma^k_{\ k}$ are modified
in the region: $0\le r\le 6\kappa^{-1}$.
Far from the dislocation line ($r\gg 12\kappa^{-1}$) the gauge theoretical 
and the classical solutions of the stress of an edge dislocation coincide.
Thus, the characteristic internal length $\kappa^{-1}$ determines the
position and the magnitude of the stress maxima.
It is interesting and important to note that the 
gauge-theoretical solutions (\ref{T_xx})-(\ref{T_zz}) agree precisely 
with the gradient solutions given by Gutkin and 
Aifantis~\cite{GA99,Gutkin00} 
(with $\kappa^{-2}=c$, $c$ is the gradient coefficient).

For the elastic strain of an edge dislocation we find
\begin{align}
&E_{xx}=-\frac{b}{4\pi(1-\nu)}\, 
\frac{y}{r^2}
\bigg\{(1-2\nu)+\frac{2x^2}{r^2}+\frac{4}{\kappa^2r^4}\big(y^2-3x^2\big)\\
&\hspace{5cm}-2\left(\frac{y^2}{r^2}-\nu\right)\kappa r K_1(\kappa r)
-\frac{2}{r^2}\big(y^2-3x^2\big) K_2(\kappa r)\bigg\},\nonumber\\
&E_{yy}=-\frac{b}{4\pi(1-\nu)}\, 
\frac{y}{r^2}
\bigg\{(1-2\nu)-\frac{2x^2}{r^2}-\frac{4}{\kappa^2r^4}\big(y^2-3x^2\big)\\
&\hspace{5cm}-2\left(\frac{x^2}{r^2}-\nu\right)\kappa r K_1(\kappa r)
+\frac{2}{r^2}\big(y^2-3x^2\big) K_2(\kappa r)\bigg\},\nonumber\\
&E_{xy}=\frac{b}{4\pi(1-\nu)}\, 
\frac{x}{r^2}
\bigg\{1-\frac{2y^2}{r^2}-\frac{4}{\kappa^2r^4}\big(x^2-3y^2\big)\\
&\hspace{5cm} -\frac{2y^2}{r^2}\,\kappa r K_1(\kappa r)
+\frac{2}{r^2}\big(x^2-3y^2\big) K_2(\kappa r)\bigg\},\nonumber
\end{align}
which is in agreement with the solution given by Gutkin and Aifantis~\cite{GA97,GA99,Gutkin00}
in the framework of strain gradient elasticity.
The plane-strain condition $E_{zz}=0$ of the classical 
dislocation theory is also valid in the dislocation core region. 
The components of the strain tensor have the
following extreme values ($\nu=0.3$):
$|E_{xx}(0,y)|\simeq 0.308\kappa \frac{ b}{4\pi(1-\nu)}$ at 
$|y|\simeq 0.922 \kappa^{-1}$,
$|E_{yy}(0,y)|\simeq 0.010 \kappa\frac{ b}{4\pi(1-\nu)}$ at 
$|y|\simeq 0.218 \kappa^{-1}$, 
$|E_{yy}(0,y)|\simeq 0.054 \kappa\frac{ b}{4\pi(1-\nu)}$ at 
$|y|\simeq 4.130 \kappa^{-1}$, 
and
$|E_{xy}(x,0)|\simeq 0.260 \kappa\frac{ b}{4\pi(1-\nu)}$ at 
$|x|\simeq 1.494 \kappa^{-1}$. 
%%\begin{figure}[t]\unitlength1cm
%%\vspace*{-1.0cm}
%%\centerline{
%%\begin{picture}(7,5)
%%\put(0.0,0.2){\epsfig{file=Tzz.eps,width=8cm}}
%%\put(4.0,0.0){$\kappa y$}
%%\put(-3.0,4.0){$E_{xx}(0,y)$, $E_{yy}(0,y)$}
%%\end{picture}
%%}
%%\caption{The stress components 
%%$\tl E {}_{xx}(0,y)\equiv\tl E {}_{yy}(0,y)$ (dashed curve),
%%$E_{xx}(0,y)$, $E_{yy}(0,y)$ near the dislocation line. 
%%The strain is given in units of $b\kappa/[4\pi(1-\nu)]$.}
%%\label{fig:strain}
%%\end{figure}
It is interesting to note that 
$E_{yy}(0,y)$ is much smaller than $E_{xx}(0,y)$ within
the core region (see also~\cite{GA99,Gutkin00}).
The dilatation $E^k_{\ k}$ reads
\begin{align}
E^k_{\ k}=-\frac{b(1-2\nu)}{2\pi(1-\nu)}\, 
\frac{y}{r^2}\Big\{1-\kappa  r K_1(\kappa r)\Big\}.
\end{align}

It is convenient to rewrite the stress and strain fields in cylindrical
coordinates.
The stress function reads in cylindrical coordinates:
\begin{align}
\label{f_edge-polar}
f=-\frac{\mu b}{2\pi(1-\nu)}\, \sin\varphi \left\{r\ln r 
+\frac{2}{\kappa^2 r}\Big(1-\kappa r K_1(\kappa r)\Big)\right\}. 
%%%%%%%+\text{const}.
\end{align}
Here the stress function is related to the stresses by the equations 
\begin{align}
\sigma_{rr}=\frac{1}{r}\,\pd_r f+\frac{1}{r^2}\,\pd^2_{\varphi\varphi} f,\quad
\sigma_{\varphi\varphi}=\pd^2_{rr} f,\quad
\sigma_{r\varphi}=-\pd_r\left(\frac{1}{r}\,\pd_\varphi f\right),\quad
\sigma_{zz}=\nu(\sigma_{rr}+\sigma_{\varphi\varphi}).
\end{align}
In this way we find the stress of a straight edge dislocation in cylindrical 
coordinates
\begin{align}
\label{T_rr}
\sigma_{rr}&=-\frac{\mu b}{2\pi(1-\nu)}\, 
\frac{\sin\varphi}{r}\left\{1-\frac{4}{\kappa^2r^2}+2 K_2(\kappa r)\right\},\\
\label{T_rp}
\sigma_{r\varphi}&=\frac{\mu b}{2\pi(1-\nu)}\, 
\frac{\cos\varphi}{r}\left\{1-\frac{4}{\kappa^2r^2}+2 K_2(\kappa r)\right\},\\
\label{T_pp}
\sigma_{\varphi\varphi}&=-\frac{\mu b}{2\pi(1-\nu)}\, 
\frac{\sin\varphi}{r}\left\{1+\frac{4}{\kappa^2 r^2}
-2 K_2(\kappa r)-2\kappa r K_1(\kappa r)\right\},\\
\label{T_zz2}
\sigma_{zz}&=-\frac{\mu b\nu }{\pi(1-\nu)}\, 
\frac{\sin\varphi}{r}\Big\{1-\kappa r K_1(\kappa r)\Big\},
\end{align}
and the trace of the stress tensor reads
\begin{align}
\label{hyd_p-polar}
\sigma^k_{\ k}=-\frac{\mu b(1+\nu)}{\pi(1-\nu)}\, 
\frac{\sin\varphi}{r}\Big\{1-\kappa  r K_1(\kappa r)\Big\}.
\end{align}
Eventually, the elastic strain of a straight edge dislocation 
is given in cylindrical coordinates as follows
\begin{align}
\label{}
E_{rr}&=-\frac{b}{4\pi(1-\nu)}\, 
\frac{\sin\varphi}{r}\left\{(1-2\nu)-\frac{4}{\kappa^2r^2}+2 K_2(\kappa r)
+2\nu\kappa r K_1(\kappa r)\right\},\\
\label{}
E_{r\varphi}&=\frac{b}{4\pi(1-\nu)}\, 
\frac{\cos\varphi}{r}\left\{1-\frac{4}{\kappa^2r^2}+2 K_2(\kappa r)\right\},\\
E_{\varphi\varphi}&=-\frac{b}{4\pi(1-\nu)}\, 
\frac{\sin\varphi}{r}\left\{(1-2\nu)+\frac{4}{\kappa^2 r^2}
-2 K_2(\kappa r)-2(1-\nu)\kappa r K_1(\kappa r)\right\}.
\end{align}
The dilatation reads
\begin{align}
E^k_{\ k}=-\frac{b(1-2\nu)}{2\pi(1-\nu)}\, 
\frac{\sin\varphi}{r}\Big\{1-\kappa  r K_1(\kappa r)\Big\}.
\end{align}

Let us now calculate the distortion of an edge dislocation.
The distortion $\beta^a_{\ i}$ is given in terms of the stress function~(\ref{f_edge1}):
\begin{align}
\label{dist-ansatz}
\beta^a_{\ i}=\frac{1}{2\mu}
\left(\begin{array}{ccc}
\pd^2_{yy}f-\nu\Delta f & -\pd^2_{xy}f +2\mu\omega& 0\\
-\pd^2_{xy}f-2\mu\omega & \pd^2_{xx}f-\nu\Delta f & 0\\
0& 0& 0
\end{array}\right),
\end{align}
where $\omega$ is used to express the antisymmetric part of the distortion,
$\omega\equiv\beta_{[xy]}$.
Eventually, $\omega$ is determined from the conditions:
\begin{align}
\label{alpha_xz}
T^x_{\ xy}&=\frac{1}{2\mu}\big(2\mu\pd_x\omega-(1-\nu)\pd_y\Delta f\big),\\
\label{alpha_yz}
T^y_{\ xy}&=\frac{1}{2\mu}\big(2\mu\pd_y\omega+(1-\nu)\pd_x\Delta f\big)\equiv 0.
\end{align}
We find for the elastic distortion of the straight dislocation
\begin{align}
\label{dist_xx}
&\beta^{x}_{\ x}=-\frac{b}{4\pi(1-\nu)}\, 
\frac{y}{r^2}
\bigg\{(1-2\nu)+\frac{2x^2}{r^2}+\frac{4}{\kappa^2r^4}\big(y^2-3x^2\big)\\
&\hspace{4cm}
-2\left(\frac{y^2}{r^2}-\nu\right)\kappa r K_1(\kappa r)
-\frac{2}{r^2}\big(y^2-3x^2\big) K_2(\kappa r)\bigg\},\nonumber\\
&\beta^{x}_{\ y}=\frac{b}{4\pi(1-\nu)}\, 
\frac{x}{r^2}
\bigg\{(3-2\nu)-\frac{2y^2}{r^2}-\frac{4}{\kappa^2r^4}\big(x^2-3y^2\big)
\\
&\hspace{4cm} 
-2\left((1-\nu)+\frac{y^2}{r^2}\right)\kappa r K_1(\kappa r)
+\frac{2}{r^2}\big(x^2-3y^2\big) K_2(\kappa r)\bigg\},\nonumber\\
&\beta^{y}_{\ x}=-\frac{b}{4\pi(1-\nu)}\, 
\frac{x}{r^2}
\bigg\{(1-2\nu)+\frac{2y^2}{r^2}+\frac{4}{\kappa^2r^4}\big(x^2-3y^2\big)\\
&\hspace{4cm}
-2\left((1-\nu)-\frac{y^2}{r^2}\right)\kappa r K_1(\kappa r)
-\frac{2}{r^2}\big(x^2-3y^2\big) K_2(\kappa r)\bigg\},\nonumber\\
\label{dist_yy}
&\beta^{y}_{\ y}=-\frac{b}{4\pi(1-\nu)}\, 
\frac{y}{r^2}
\bigg\{(1-2\nu)-\frac{2x^2}{r^2}-\frac{4}{\kappa^2r^4}\big(y^2-3x^2\big)\\
&\hspace{4cm}
-2\left(\frac{x^2}{r^2}-\nu\right)\kappa r K_1(\kappa r)
+\frac{2}{r^2}\big(y^2-3x^2\big) K_2(\kappa r)\bigg\}\nonumber,
\end{align}
and for the rotation
\begin{align}
\label{rot_z}
\omega_z\equiv-\omega=-\frac{b}{2\pi}\,\frac{x}{r^2}\Big\{1-\kappa r K_1(\kappa r)\Big\}.
\end{align}
Eq.~(\ref{rot_z}) is in agreement with the rotation vector calculated in the
linear theory of dislocations in the Cosserat continuum~\cite{Nowacki73}.
The far fields of~(\ref{dist_xx})--(\ref{rot_z}) are identical
to the classical ones given in~\cite{deWit73b}.

\begin{figure}[t]\unitlength1cm
%%\vspace*{-0.5cm}
\centerline{
(a)
\begin{picture}(9,6)
\put(0.0,0.2){\epsfig{file=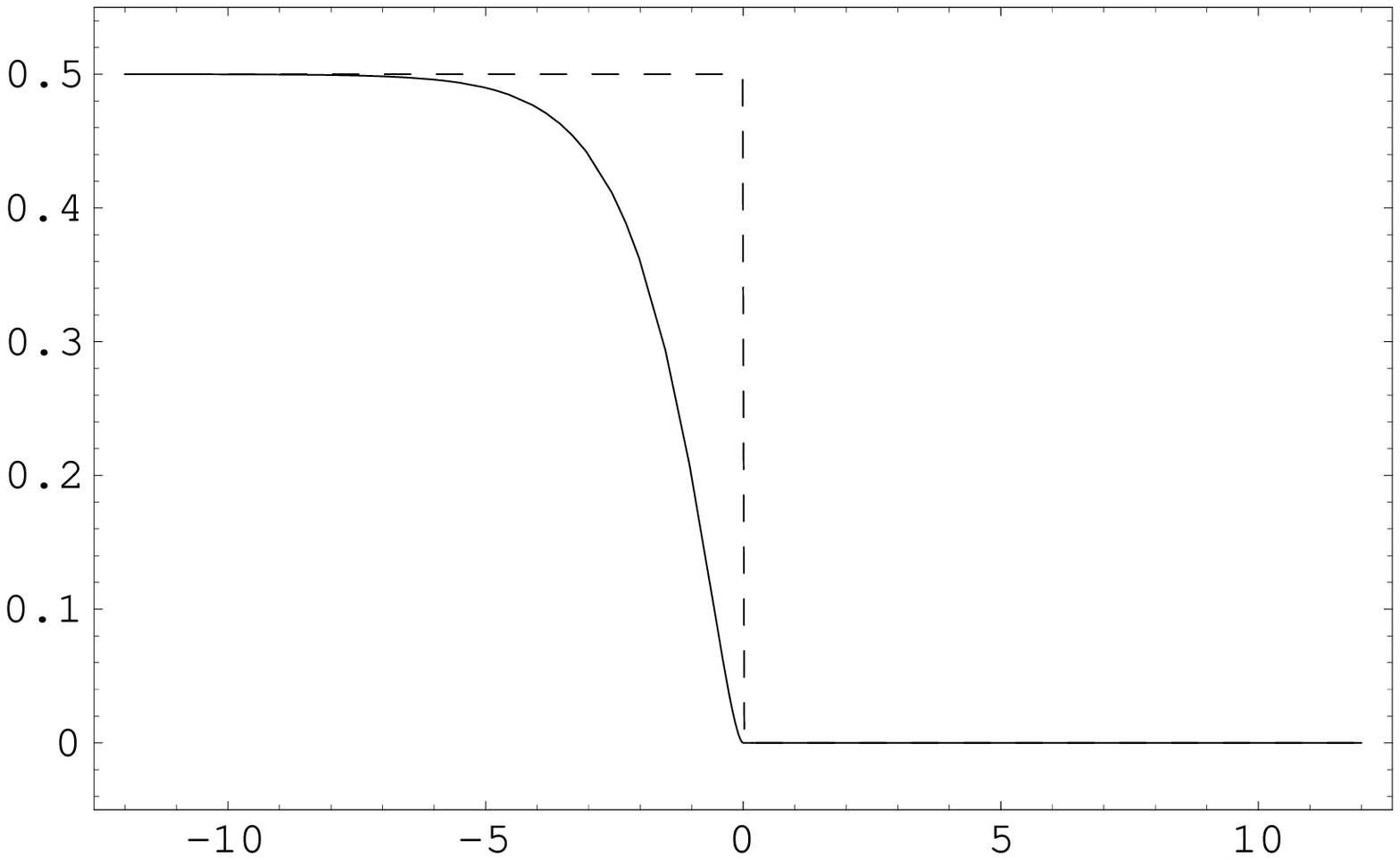,width=9cm}}
\put(4.5,0.0){$\kappa x$}
\put(-1.7,4.5){$u^x(x,0)/b$}
\end{picture}
}
%%\caption{Displacement field $u^x/b$.}
%%\label{fig:u_x}
%%\end{figure}
%%\begin{figure}[t]\unitlength1cm
%\vspace*{-1.0cm}
\centerline{
(b)
\begin{picture}(9,6)
\put(0.0,0.2){\epsfig{file=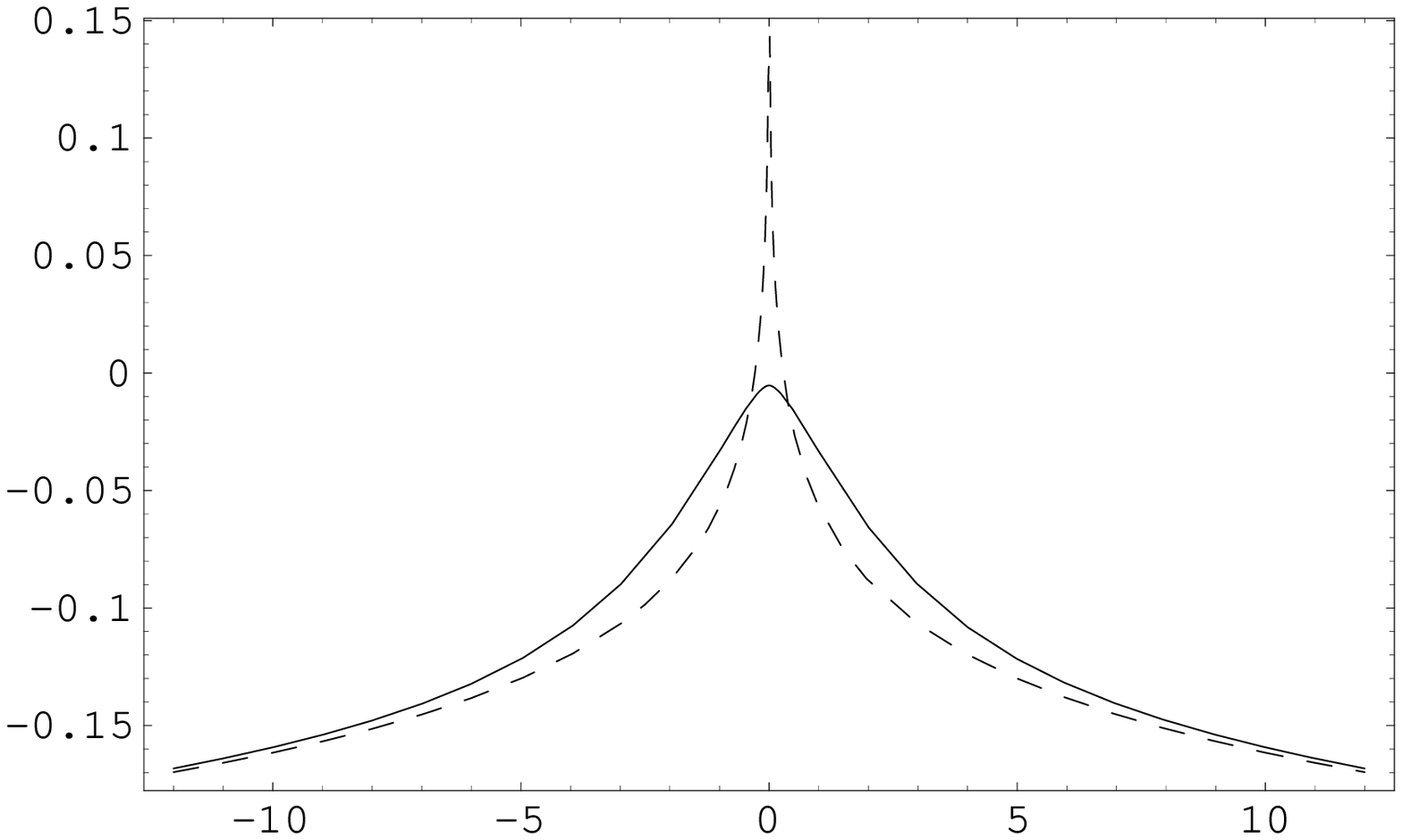,width=9.0cm}}
\put(4.5,0.0){$\kappa x$}
\put(-1.7,4.5){$u^y(x,0)/b$}
\end{picture}
}
\caption{Effective displacement fields ($\nu=0.3$): $u^x(x,y\rightarrow+0)/b$, (b) $u^y(x,0)/b$.
The dashed curves represent the classical solution.}
\label{fig:u_x}
\end{figure}

We now rewrite the distortions~(\ref{dist_xx})--(\ref{dist_yy})
as follows
\begin{align}
\label{dist_x}
\beta^{x}&=\frac{b}{2\pi}
\bigg\{\bigg(\big(1-\kappa r K_1(\kappa r)\big)
+\frac{1}{2(1-\nu)}\,
\cos 2\varphi\left(1-\frac{4}{\kappa^2r^2}+2 K_2(\kappa r)\right)\bigg)
\d\varphi\nonumber\\
&\hspace{3cm}+\frac{1}{2(1-\nu)}\,
\sin 2\varphi\left(\frac{4}{\kappa^2r^3}-\kappa K_1(\kappa r)-\frac{2}{r} K_2(\kappa r)\right)
\d r\bigg\}\nonumber\\
&\equiv \d u^x +\phi^x.
\end{align}
We may interpret the field 
\begin{align}
\phi^x=-\frac{b\kappa^2}{2\pi}\,\varphi r K_0(\kappa r)\,\d r
\end{align}
as the proper incompatible part (negative plastic distortion). 
This proper incompatible distortion is exactly the same as 
$\phi^z$ of a screw dislocation (see~\cite{Lazar02,Lazar02b}).
In components it reads
\begin{align}
\label{plastic-dist}
\phi^x_{\ x}=-\frac{b\kappa^2}{2\pi}\,x\varphi K_0(\kappa r),\qquad
\phi^x_{\ y}=-\frac{b\kappa^2}{2\pi}\,y\varphi K_0(\kappa r).
\end{align}
The appearance of such  proper incompatible distortion is a typical 
result of the gauge theory of dislocations.
When $\kappa^{-1}\rightarrow 0$, the incompatible
distortion~(\ref{plastic-dist}) converts to 
$\phi^x_{\ x}=-b\,x\varphi\,\delta(r)$ and $\phi^x_{\ y}=-b\,y\varphi\,\delta(r)$.
%%%%It has an extremum at about $r\simeq 0.60\kappa^{-1}$ with 
%%%%$|\hat{\phi}^z_{\rm max}|\simeq 0.466 \frac{b}{2\pi} \kappa$ 
%%%%(see Fig.~\ref{fig:gauge}).
The compatible part of Eq.~(\ref{dist_x}) is given in terms of the
displacement field
\begin{align}
\label{u_x}
u^x=\frac{b}{2\pi}\bigg\{\big(1-\kappa r K_1(\kappa r)\big)\varphi
+\frac{1}{4(1-\nu)}\, \sin 2\varphi\left(1-\frac{4}{\kappa^2r^2}+2 K_2(\kappa r)\right)\bigg\}.
%%%%%+\text{const}
\end{align}
It is a modified displacement field (see Fig.~\ref{fig:u_x}a). 
We used $\varphi=\arctan (y/x)$ as a multi-valued field.
Note that the first part in (\ref{u_x}) is different from the corresponding
one given in gradient elasticity by Gutkin and Aifantis~\cite{GA97,GA99}. 
One reason is that they used a compatible elastic distortion.
%%%%%%This $u^x$ is multi-valued and has no singularity.
It is important to bear in mind that $u^x$ is not a simple $\arctan$ function 
as in the Volterra model of a dislocation. 
Additionally, we find from the distortion
\begin{align}
\label{dist_y}
\beta^{y}&=-\frac{b}{4\pi(1-\nu)}
\bigg\{
\sin 2\varphi\left(1-\frac{4}{\kappa^2r^2}+2 K_2(\kappa r)\right)
\d\varphi\nonumber\\
&\qquad
-\bigg((1-2\nu)\left(\frac{1}{r}-\kappa  K_1(\kappa r)\right)
+\cos 2\varphi\left(\frac{4}{\kappa^2r^3}-\kappa K_1(\kappa r)-\frac{2}{r} K_2(\kappa r)\right)\bigg)
\d r\bigg\}\nonumber\\
&\equiv \d u^y +\phi^y ,
\end{align}
the incompatible part
\begin{align}
\phi^y=0,
\end{align}
and a single-valued displacement field
\begin{align}
\label{u_y}
u^y=-\frac{b}{8\pi(1-\nu)}\bigg\{2(1-2\nu)\big(\ln r+ K_0(\kappa r)\big)
+\cos 2\varphi\left(1-\frac{4}{\kappa^2r^2}+2 K_2(\kappa r)\right)\bigg\}
%%%%%+\text{const}
\end{align}
which has no singularity (see  Fig.~\ref{fig:u_x}b). 
This means that $\beta^y$ is a proper compatible distortion.
That is the reason why the field $u^y$ agrees precisely with the
corresponding formula in gradient elasticity (see~\cite{GA97,GA99}).
The displacement field
$u^y$ differs from the classical one in the region $0\le r\le 12\kappa^{-1}$.
In this framework the displacement fields~(\ref{u_x}) and
(\ref{u_y}) have no singularity. 
%%%%We have calculated the core configuration of an edge dislocation in a
%%%simple cubic lattice for two different choices of the coefficient $\kappa^{-1}$ 
%%%(see Fig.~\ref{fig:edge-core}). 
%%%\begin{figure}[t]\unitlength1cm
%%%\centerline{
%%%\epsfig{figure=edge-GA.eps,width=6.0cm}
%%\put(-6.3,3.0){$\kappa y$}
%%%\put(-6.0,-0.5){$\text{(a)}$}
%%%\hspace*{0.5cm}
%%%\put(0,-0.5){$\text{(b)}$}
%%%\epsfig{figure=edge-Er.eps,width=6.0cm}
%%%}
%%%\caption{Perfect edge dislocation in a simple cubic lattice:
%%%(a) $\kappa^{-1}=0.25a$,
%%%(b) $\kappa^{-1}=0.399a$. The position $x=1/2$ and $y=1/2$ is used as the 
%%%centre of the dislocation.}
%%%\label{fig:edge-core}
%%%\end{figure}
%%%
\begin{figure}[t]\unitlength1cm
\centerline{
\begin{picture}(9,6)
\put(0.0,0.2){\epsfig{file=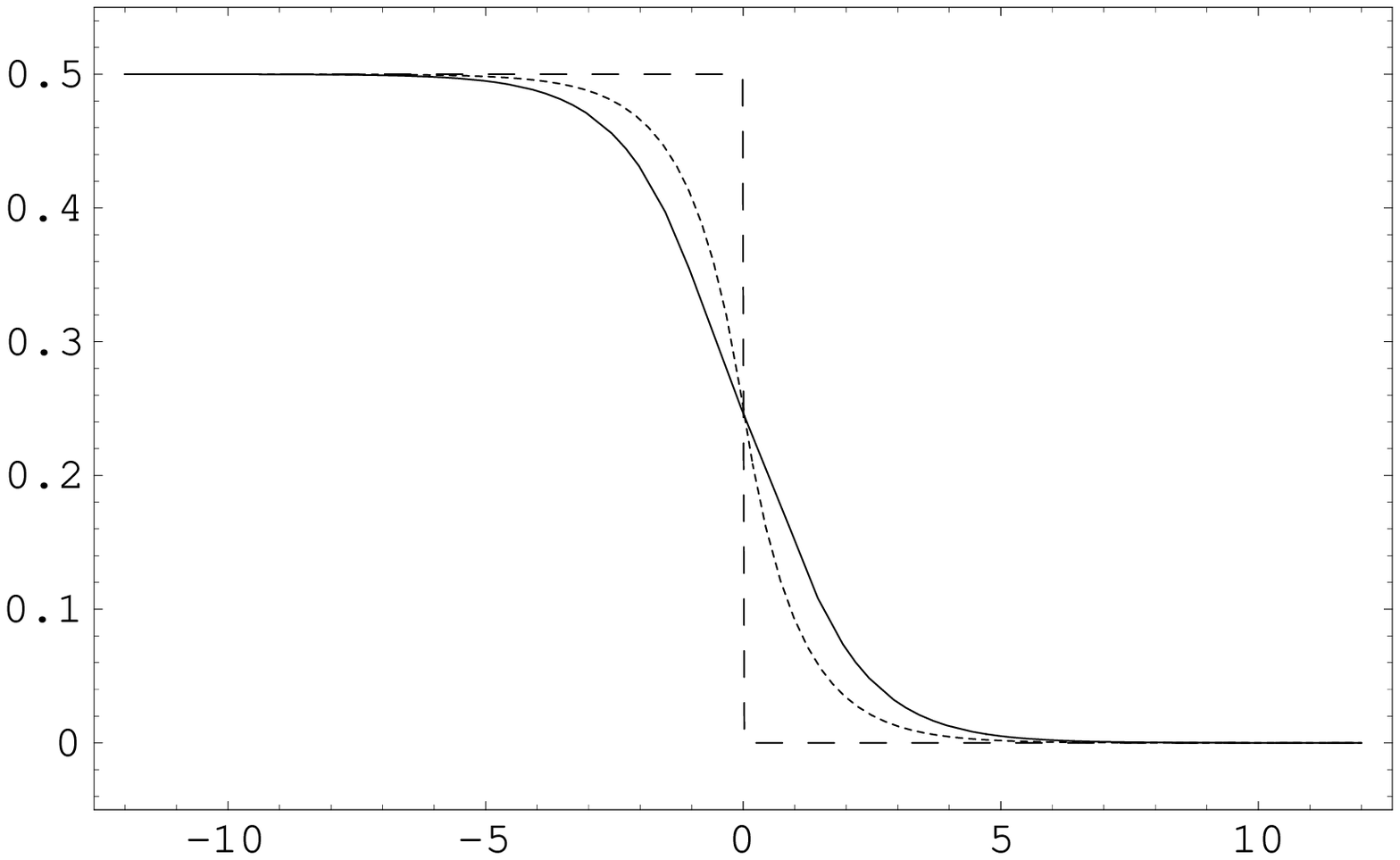,width=9cm}}
\put(4.5,0.0){$\kappa x$}
\put(-1.7,4.5){$u^x(x,0)/b$}
\end{picture}
}
%%%\centerline{\epsfxsize=9cm\epsffile{burger-scr.eps}}
\caption{Displacement $u^x(x,y\rightarrow +0)/b$ 
near the line of edge dislocation within gauge model (solid),
the gradient model (small dashed) and the classical elasticity (dashed).
We used $\kappa^2=1/c$.}
\label{fig:u-total}
\end{figure}
%%%%In Fig.~\ref{fig:edge-core}a we have used the value of $\kappa^{-1}=0.25 a$
%%%%which was proposed by Altan and Aifantis~\cite{AA92,Aifantis94,AA97}. 
%%%%The value $\kappa^{-1}=0.399 a$ proposed by Eringen~\cite{Eringen83,Eringen85}
%%%%is used in Fig.~\ref{fig:edge-core}b. 
%%%%It can be seen that the dislocation core in Fig.~\ref{fig:edge-core}a 
%%%%%has a  symmetrical shape. The added half plane ends at the centre of the dislocation.
%%%%The core shape is changed to an unsymmetrical configuration in  Fig.~\ref{fig:edge-core}b 
%%%%and its centre is shifted slightly. 
The far fields of the displacements~(\ref{u_x}) and (\ref{u_y}) are identical
to the classical ones given in~\cite{HL,Read,Teodosiu}.
It can be seen that the solution~(\ref{u_x}) leads to an asymmetric 
smoothing of the displacement profile (see Fig.~\ref{fig:u_x}a). 
In fact, it is asymmetric around the dislocation line.
Note that the incompatible distortion~(\ref{plastic-dist}) 
also has an asymmetric form. 
These asymmetries would lead to an unexpected asymmetry of the 
dislocation core. 
From the mathematical point of view, the asymmetry is induced by the 
decomposition of the elastic distortion~(\ref{dist2}) into compatible and 
incompatible parts.
Keep in mind that such a decomposition is gauge invariant (see Eq.~(\ref{gauge-trans}))
and we have a local translation as an additional degree of freedom. 
This gauge invariance can be used to improve the situation.
So far we have considered $\varphi$ as a multi-valued field. 
On the other hand, in the defect theory~\cite{GA96,GA97,GA99,Gutkin00,deWit73b} $\varphi$ 
is usually used as single-valued and discontinuous function. 
It is made unique by cutting the half-plane $y=0$ at $x<0$ and 
assuming $\varphi$ to jump from $\pi$ to $-\pi$ when crossing the cut.
If we use the single-valued discontinuous form for $\varphi$ and 
the local gauge transformation~(\ref{gauge-trans}) 
with the gauge function
\begin{align}
\tau^x=\frac{b}{2\pi}\, {\mathrm{sign}} (y)\,\frac{\pi}{2}\,\kappa r K_1(\kappa r),
\end{align}
we obtain for the improved displacement field
\begin{align}
\label{u_x2}
u^x&=\frac{b}{2\pi}\bigg\{\varphi\big(1-\kappa r K_1(\kappa r)\big)
+\frac{\pi}{2}\, {\mathrm{sign}} (y)\,\kappa r K_1(\kappa r)\nonumber\\
&\qquad\qquad\qquad
+\frac{1}{2(1-\nu)}\, \frac{xy}{r^2}\left(1-\frac{4}{\kappa^2r^2}+2 K_2(\kappa r)\right)\bigg\},
\end{align}
and for the proper incompatible distortion
\begin{align}
\label{plastic-dist2}
&\phi^x_{\ x}=-\frac{b}{2\pi}\,\kappa^2 x K_0(\kappa r)
\left(\varphi-\frac{\pi}{2}\, {\mathrm{sign}}(y)\right),\\
&\phi^x_{\ y}=-\frac{b}{2\pi}
\label{plastic-dist2-y}
\left\{\kappa^2 y K_0(\kappa r)\left(\varphi-\frac{\pi}{2}\, {\mathrm{sign}}(y)\right)
+\pi\delta(y) \Big(1-{\mathrm{sign}}(x)\big[1-\kappa r K_1(\kappa r)\big]\Big)\right\}.
\end{align}
Now the displacements and the incompatible distortions are symmetric 
in the core region around the dislocation line\footnote{
Note that in the case of a screw dislocation the symmetrical displacement 
and the corresponding incompatible distortion read, respectively,
\begin{align*}
u^z=\frac{b}{2\pi}\left\{\varphi\big(1-\kappa r K_1(\kappa r)\big)
+\frac{\pi}{2}\, {\mathrm{sign}} (y) \kappa r K_1(\kappa r)\right\},
\end{align*}
and
\begin{align}
%%%\label{plastic-dist2}
&\phi^z_{\ x}=-\frac{b}{2\pi}\,\kappa^2 x K_0(\kappa r)
\left(\varphi-\frac{\pi}{2}\, {\mathrm{sign}}(y)\right),\nonumber\\
&\phi^z_{\ y}=-\frac{b}{2\pi}
\left\{\kappa^2 y K_0(\kappa r)\left(\varphi-\frac{\pi}{2}\, {\mathrm{sign}}(y)\right)
+\pi\delta(y)\Big(1-{\mathrm{sign}}(x)
\big[1-\kappa r K_1(\kappa r)\big]\Big)\right\}.\nonumber
\end{align}
}.
It is interesting to compare the displacement field~(\ref{u_x2}) with 
the displacement calculated by Gutkin and Aifantis. 
When $y\rightarrow 0$,
%%% transforms into 
%%%\begin{align}
%%%u^x(x,y\rightarrow 0)=
%%%\frac{b}{2\pi}\,{\mathrm{sign}}(y)\,\frac{\pi}{2}
%%%\Big\{1-{\mathrm{sign}}(x)\,\big(1-\kappa |x| K_1(\kappa |x|)\big)\Big\}.
%%\end{align}
the Bessel function terms in ~(\ref{u_x2})
lead to the symmetric smoothing of the displacement
profile in contrast to the abrupt jump occuring in the profile of the classical
solution (see Fig.~\ref{fig:u-total}).
When $y\rightarrow 0$, the displacement $u^x(x,y)$ calculated
by Gutkin and Aifantis~\cite{GA97,Gutkin00} within the gradient model has the explicit form
\begin{align}
\label{u-grad}
u^x(x,y\rightarrow 0)=
\frac{b}{2\pi}\,{\mathrm{sign}}(y)\,\frac{\pi}{2}
\left\{1-{\mathrm{sign}}(x)\left(1-\e^{-|x|/\sqrt{c}}\right)\right\}.
\end{align}
The exponential term which appears in the gradient solution (\ref{u-grad})
leads to the smoothing of the displacement profile 
(see Fig.~\ref{fig:u-total}). 
The size of the transition zone is approximately $12 \kappa^{-1}$
which gives the value $r_c\simeq 6\kappa^{-1}$ for the core radius.
Consequently, in the gauge theory of dislocations and in gradient elasticity 
the dislocation core appears naturally. 
Therefore, these 
displacement fields may be used to model the dislocation core. 
In this way, one can compare the calculated dislocation core
with HRTEM micrographs and related computer simulations of the core region. 
It would be very interesting to check a possible material dependence 
of $\kappa^{-1}$.

Finally, the effective Burgers vector can be calculated as
\begin{align}
b^x(r)=\oint_\gamma\beta^{x}
      =b\Big\{1-\kappa r K_1(\kappa r)\Big\}.
\end{align}
\begin{figure}[t]\unitlength1cm
\centerline{
\begin{picture}(9,6)
\put(0.0,0.2){\epsfig{file=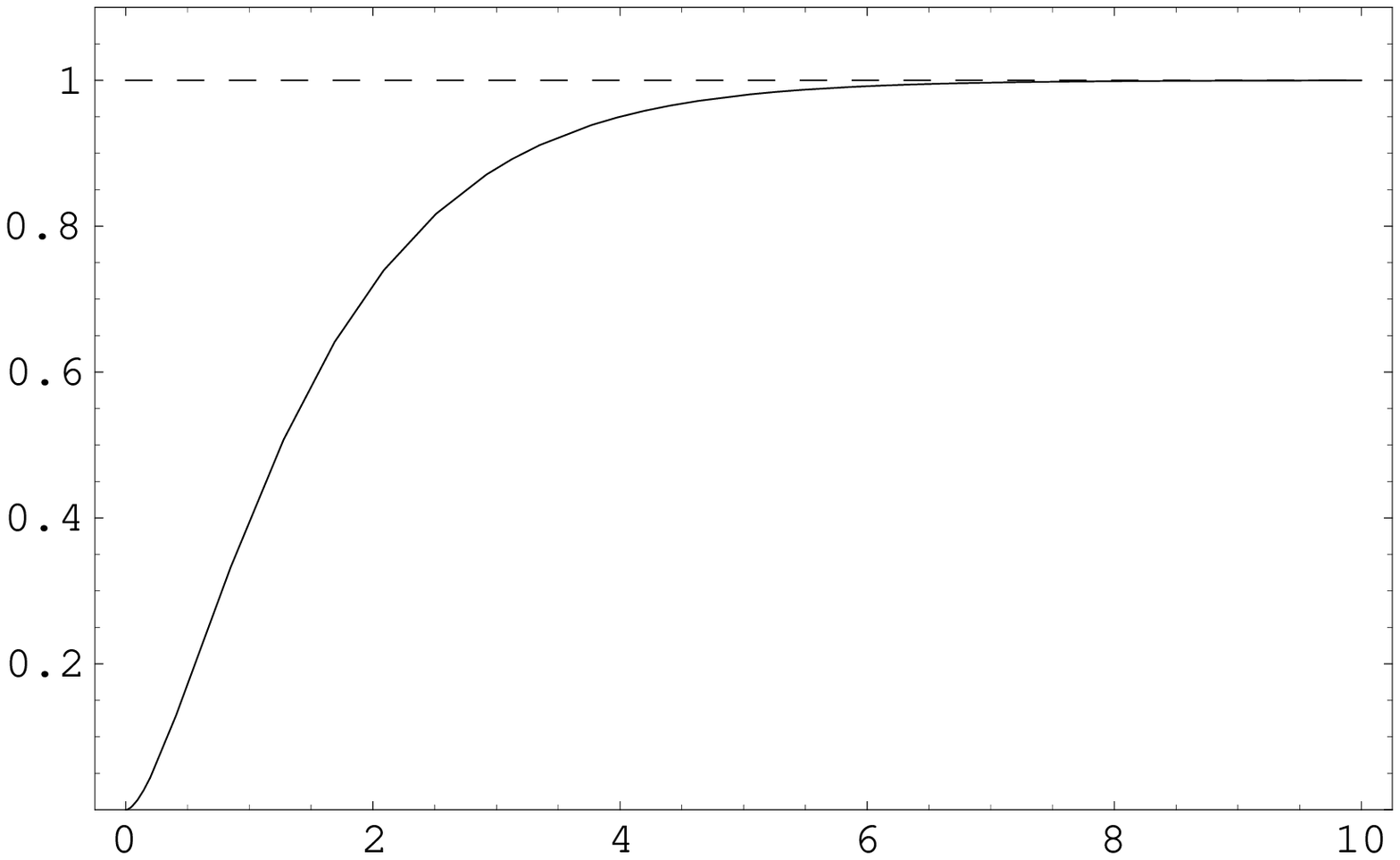,width=9cm}}
\put(4.5,0.0){$\kappa r$}
\put(-1.7,4.5){$b^x(r)/b$}
\end{picture}
}
%%%\centerline{\epsfxsize=9cm\epsffile{burger-scr.eps}}
\caption{Effective Burgers vector $b^x(r)/b$ (solid).}
\label{fig:burger}
\end{figure}
This effective Burgers vector $b^x(r)$ differs from the constant Burgers 
vector $b$ in the region $0\le r\le 6\kappa^{-1}$ (see Fig.~\ref{fig:burger}). 
For the value of $\kappa^{-1}=0.25a$ the core radius is 
$r_c=1.5a$ and for the value of $\kappa^{-1}=0.399a$ the core radius is 
$r_c=2.4a$. 
Note that the effective Burgers vector $b^x(r)$ of an edge dislocation
has the same form as the effective Burgers vector $b^z(r)$ of a screw 
dislocation which is obtained in~\cite{Lazar02,Malyshev00,VS88,Edelen96}.

\subsection{The dislocation density, moment stress and core energy}
The proper incompatible part of the elastic distortion gives rise 
to a localized torsion and moment stress tensor.
We find for the torsion
\begin{align}
T^x_{\ xy}=-T^x_{\ yx}=\frac{b\kappa^2}{2\pi}\, K_0(\kappa r)
\end{align}
and the dislocation density of an edge dislocation
\begin{align}
\label{disl-den}
\alpha^{xz}=\frac{b\kappa^2}{2\pi}\,K_0(\kappa r),
\end{align}
which satisfies the translational Bianchi identity:
$\pd_i \alpha^{ai}=0$.
On can see that the dislocation density is short-reaching.
It is interesting to note that the torsion $T^x_{\ xy}$ of an edge dislocation
has the same form as the torsion $T^z_{\ xy}$ of a screw 
dislocation which is given in~\cite{Lazar02,Lazar02b,Malyshev00,VS88,Edelen96}.
In the limit as $\kappa^{-1}\rightarrow 0$, the gauge theoretical 
result~(\ref{disl-den}) converts to the classical dislocation density
$\alpha^{xz}=b\,\delta(r)$.
Additionally, the dislocation density~(\ref{disl-den}) agrees with 
Eringen's two-dimensional nonlocal modulus (nonlocal kernel) used
in~\cite{Eringen83,AE83,Eringen85,Eringen87}.
Consequently, the dislocation density tensor is Green's function of the
Helmholtz equation:
\begin{align}
\label{Helmholtz-torsion}
\left(1-\kappa^{-2}\Delta\right)\alpha^{xz}(r)=b\,\delta(r).
\end{align}
This torsion gives rise to a localized moment stress.
The localized moment stress of bending type is given by
the help of~(\ref{moment-2}) as
\begin{align}
\label{moment-stress}
H_{xz}=\frac{\mu b}{\pi(1-\nu)}\, K_0(\kappa r),\qquad
H_{zx}=-\frac{\mu b \nu}{\pi(1-\nu)}\, K_0(\kappa r).
\end{align}
This expression is very close to the moment stress of an edge dislocation
given in~\cite{HK65}.
By means of the dislocation density and the moments stress
we are able to calculate the core energy of an edge dislocation.
The dislocation core energy density is given by
\begin{align}
V_{\text{core}}=\frac{1}{2}\,\alpha^{ai}H_{ai}
        =\frac{\mu b^2\kappa^2}{4\pi^2(1-\nu)}\, K_0(\kappa r)^2.
\end{align}
Obviously, the moment stress $H_{zx}$ gives no contribution to the
dislocation core energy.
We obtain for the dislocation core energy per unit length
of the straight edge dislocation
\begin{align}
E_{\text{core}}&=\int_0^{2\pi}\d\varphi\int_0^R r\, V_{\text{core}}\d r\nonumber\\
               &=\frac{\mu b^2\kappa^2}{4\pi(1-\nu)}\, 
                r^2\Big\{K_0(\kappa r)^2-K_1(\kappa r)^2\Big\}\Big|_0^R\nonumber\\
               &=\frac{\mu b^2}{4\pi(1-\nu)}\Big\{1 +
                \kappa^2 R^2\big(K_0(\kappa R)^2-K_1(\kappa R)^2\big)\Big\},            
\end{align}
where $R$ denotes the size of the solid body. 
In the limit $R\rightarrow\infty$, the core energy of the straight edge dislocation, 
\begin{align}
E_{\text{core}}=\frac{\mu b^2}{4\pi(1-\nu)},
\end{align}
coincides with the core energy that is calculated for 
the edge dislocation in the Peierls-Nabarro model~\cite{HL}.
The core energy is different from the constant value in the region
$0\le \kappa R\le 3$.

The far-reaching rotation gradients read
\begin{align}
\label{curv-k}
&k^x_{\ xy}=-k^x_{\ yx}=-\frac{b}{2\pi r^4}\Big\{\big(x^2-y^2\big)\big(1-\kappa r K_1(\kappa r)\big)
                -\kappa^2 x^2 r^2 K_0(\kappa r)\Big\},\nonumber \\
&k^y_{\ xy}=-k^y_{\ yx}=-\frac{b}{2\pi r^4}\, xy \Big\{2\big(1-\kappa r K_1(\kappa r)\big)
                -\kappa^2 r^2 K_0(\kappa r)\Big\}.
\end{align}
The form of these rotation gradients is 
in agreement with the expressions calculated within the theory of Cosserat 
media (see~\cite{Kessel70,Nowacki74}).

\subsection{The straight edge dislocation in a cylinder}
Now we consider a straight edge dislocation in a cylinder of
finite outer radius $R$ with an outer boundary free from external forces.
An advantage in our framework is that we do not need an
inner radius because the stress and strain fields are zero at $r=0$ in quite
natural way.
If the outer boundary is to be free, it is necessary to add to the stress
function (\ref{Airy1}) an additional stress function $\chi'=A\,C y r^2$.
Then the modified  stress function is of the form
\begin{align}
\label{f_edge-polar-cyl}
f=-\frac{\mu b}{2\pi(1-\nu)}\, \sin\varphi \left\{r\ln r 
+\frac{2}{\kappa^2 r}\Big(1-\kappa r K_1(\kappa r)-4C r^2\Big)-C r^3\right\},
%%%%%%+\text{const},
\end{align}
where the constant $C$ is determined by the semi-classical boundary condition 
that the stresses
$\sigma_{rr}$ and $\sigma_{r\varphi}$ should be zero at $r=R$.
In this way we find the stress of the edge dislocation in cylindrical 
coordinates
\begin{align}
\label{T_rr-cyl}
\sigma_{rr}&=-\frac{\mu b}{2\pi(1-\nu)}\, 
\frac{\sin\varphi}{r}\left\{1-\frac{4}{\kappa^2r^2}+2 K_2(\kappa r)-2C r^2\right\},\\
\label{T_rp-cyl}
\sigma_{r\varphi}&=\frac{\mu b}{2\pi(1-\nu)}\, 
\frac{\cos\varphi}{r}\left\{1-\frac{4}{\kappa^2r^2}+2 K_2(\kappa r)-2C r^2\right\},\\
\label{T_pp-cyl}
\sigma_{\varphi\varphi}&=-\frac{\mu b}{2\pi(1-\nu)}\, 
\frac{\sin\varphi}{r}\left\{1+\frac{4}{\kappa^2 r^2}
-2 K_2(\kappa r)-2\kappa r K_1(\kappa r)-6C r^2\right\},\\
\label{T_zz2-cyl}
\sigma_{zz}&=-\frac{\mu b\nu }{\pi(1-\nu)}\, 
\frac{\sin\varphi}{r}\Big\{1-\kappa r K_1(\kappa r)-4C r^2\Big\},
\end{align}
with 
\begin{align}
C=\frac{1}{2R^2}\left\{1-\frac{4}{\kappa^2 R^2}+2K_2(\kappa R)\right\}.
\end{align}
The corresponding trace is given by
\begin{align}
\label{hyd_p-cyl}
\sigma^k_{\ k}=-\frac{\mu b(1+\nu)}{\pi(1-\nu)}\, 
\frac{\sin\varphi}{r}\Big\{1-\kappa  r K_1(\kappa r)-4 C r^2\Big\}.
\end{align}
The elastic strain of this edge dislocation 
in a cylinder is
\begin{align}
\label{}
E_{rr}&=-\frac{b}{4\pi(1-\nu)}\, 
\frac{\sin\varphi}{r}\left\{(1-2\nu)-\frac{4}{\kappa^2r^2}+2 K_2(\kappa r)
+2\nu\kappa r K_1(\kappa r)-(2-8\nu)C r^2\right\},\\
\label{}
E_{r\varphi}&=\frac{b}{4\pi(1-\nu)}\, 
\frac{\cos\varphi}{r}\left\{1-\frac{4}{\kappa^2r^2}+2 K_2(\kappa r)-2C r^2\right\},\\
E_{\varphi\varphi}&=-\frac{b}{4\pi(1-\nu)}\, 
\frac{\sin\varphi}{r}\bigg\{(1-2\nu)+\frac{4}{\kappa^2 r^2}
-2 K_2(\kappa r)-2(1-\nu)\kappa r K_1(\kappa r)\nonumber\\
&\hspace{8cm}
-(6-8\nu)C r^2\bigg\}.
\end{align}
The corresponding dilatation field reads
\begin{align}
E^k_{\ k}=-\frac{b(1-2\nu)}{2\pi(1-\nu)}\, 
\frac{\sin\varphi}{r}\Big\{1-\kappa  r K_1(\kappa r)-4C r^2\Big\}.
\end{align}

From the condition~(\ref{alpha_yz}) we find for the
rotation of an edge dislocation in a cylinder
\begin{align}
\label{rot_z-cyl}
\omega_z\equiv-\omega=-\frac{b}{2\pi}\,\frac{\cos\varphi}{r}\Big\{1-\kappa r K_1(\kappa r)+4Cr^2\Big\},
\end{align}
which gives an additional constant contribution to the bend-twist tensor 
$k^x_{\ xy}$ in Eq.~(\ref{curv-k}).
The dislocation density of the edge dislocation in a cylinder is 
determined by the condition~(\ref{alpha_xz}) and it has the same form 
as the dislocation in an infinitely extended body~(\ref{disl-den}). 

For convenience we give the displacement components in 
Cartesian coordinates
\begin{align}
\label{u_x-cyl}
u^x&=\frac{b}{2\pi}\bigg\{\varphi\big(1-\kappa r K_1(\kappa r)\big)
+\frac{\pi}{2}\, {\mathrm{sign}}\, (y)\kappa r K_1(\kappa r)\nonumber\\
&\qquad\qquad
+\frac{1}{2(1-\nu)}\, \frac{xy}{r^2}\left(1-\frac{4}{\kappa^2r^2}+2 K_2(\kappa r)\right)
+\frac{C(3-4\nu)}{(1-\nu)}\, xy\bigg\},\\
\label{u_y-cyl}
u^y&=-\frac{b}{4\pi(1-\nu)}\bigg\{(1-2\nu)\big(\ln r+ K_0(\kappa r)\big)
+\frac{x^2-y^2}{2r^2}\left(1-\frac{4}{\kappa^2r^2}+2 K_2(\kappa r)\right)\nonumber\\
&\qquad\qquad
+C \Big(x^2(5-4\nu)-y^2(1-4\nu)\Big)\bigg\}.
\end{align}
Note that 
this boundary-value problem of the edge dislocation
may be applied to describe dislocation behaviour in nanoparticles
and nanowires.

\section{The relation to the nonlocal theory of dislocations}
\setcounter{equation}{0}
Let me now give the relation to Eringen's so-called nonlocal theory of 
dislocations~\cite{Eringen83,Eringen85,Eringen87}.
It includes the effect of long range interatomic forces so that it can
be used as a continuum model of the atomic lattice dynamics.
Using Green's function of the Helmholtz equation~(\ref{Helmholtz-torsion}) 
we may solve the field equation for every component of the stress 
field~(\ref{stress-fe}) by the help of the convolution integral:
\begin{align}
\sigma_{ij}(r)=\int_V G(r-r')\,\tl\sigma {}_{ij}(r')\, \d v(r')
\end{align}
with
\begin{align}
\label{green}
G(r-r')&\equiv\frac{1}{b}\,\alpha^{xz}(r-r')
        =\frac{\kappa^2}{2\pi}\,K_0(\kappa (r-r')).
%%%\text{for an edge dislocation} \\
%%%G(r-r')&\equiv\frac{1}{b}\,\alpha^z_{\ z}(r-r')
%%      =\frac{\kappa^2}{2\pi}\,K_0(\kappa (r-r'))\quad
%%%\text{for a screw dislocation}\nonumber.
\end{align}
In this way we deduce Eringen's so-called nonlocal constitutive relation
for a linear homogeneous, isotropic solid.
Therefore, 
the gauge theoretical description of dislocations in this paper 
is close to the nonlocal elasticity of 
dislocations~\cite{Eringen83,Eringen85,Eringen87}, if one uses the special 
two-dimensional Green's function~(\ref{green}). 
The form of the nonlocal kernel (dislocation density)
is fixed by the incompatible distortion. 
Finally, we may conclude that the gauge theory of dislocation 
contains a weak nonlocality due to the moment stresses in the dislocation
core region. This observation is in agreement with Kr{\"o}ner's 
opinion that the introduction of moment stresses represents the first step
in a transition from local to a nonlocal elasticity 
theory~\cite{HK65,Kroener63,Kroener66,Kroener67}.

The characteristic
internal length in these theories is $\kappa^{-1}$. This length may be
selected to be proportional to the lattice parameter $a$ 
for a single crystal, i.e.
\begin{align}
\kappa^{-1}=e_0\, a,
\end{align}
where $e_0$ is a non-dimensional constant or material function 
which can be determined by one experiment~\cite{Eringen85}.
For $e_0=0$ we have classical elasticity and there are no effects
of moment stresses.
In this limit the nonlocal kernel must revert to the Dirac delta measure.
In~\cite{Eringen83,Eringen85}
the choice $e_0=0.399$ and in~\cite{AA92,Aifantis94,AA97} the choice 
$e_0=0.25$ are proposed (see also~\cite{Gutkin00,Lazar02,Lazar02b}).

Consequently, we can conclude that the gauge theoretical solution of the
stress field for an edge dislocation given in this paper is also 
a solution in nonlocal elasticity.
However, a difference to our approach is that Eringen's nonlocal theory works
with compatible strain instead of incompatible strain and the strains and 
displacements have the classical form with singularities at the
dislocation line.

\section{Discussion and Conclusion}
The translational gauge theory of dislocations has been employed to consider 
the straight edge dislocation in an infinitely extended body and in a cylinder. 
This translational gauge theory unifies elasticity and 
plasticity to elastoplasticity in field theoretical way.
In this framework, elastoplasticity is a theory with 
force stresses and moment stresses. 
The size of this moment stress can be calculated even from linearized 
elastoplasticity theory.
Using the stress function method, 
exact analytical solutions for the 
displacements, strain and stress fields have been derived which 
have not any artificial divergency at the dislocation core 
and the corresponding
far fields agree with the classical ones for the edge dislocation.
The strain and stress fields of an edge dislocation in an infinitely 
extended body are in exact agreement with those obtained by
Gutkin and Aifantis~\cite{GA97,GA99} through an analysis of the edge 
dislocation in the theory of gradient elasticity. 
Nevertheless, the gauge theoretical and the gradient theoretical
approaches are different and they give slightly different results,
e.g. the displacement field $u^x$ due to the incompatible distortion 
in the gauge theory of dislocations. On the other hand, the strain gradient
elasticity used a purely compatible distortion. 
However, the difference cannot be too big and the numerical values 
should be of the same order.
The dislocation density of an edge dislocation obtained in the
gauge theoretical framework is in agreement with the nonlocal kernel 
given by Ari and Eringen~\cite{Eringen83,AE83,Eringen85,Eringen87}.
From that point of view the nonlocal elasticity theory approach is close to ours. 
But, on the other hand, Eringen's nonlocal elasticity contains compatible
strains only and the strain and displacement fields have the same
form as in a classical elasticity.
In the gauge theoretical analysis the dislocation core arises naturally.
We have calculated the localized moment stress which is the
response to dislocation density and the far-reaching 
rotation gradient (bend-twist tensor).

The dislocation model used in this paper contains, in general, 10 material
constants. The total Lagrangian of this model is given in terms of 
the distortion, the dislocation density (torsion tensor) and deWit's 
bend-twist tensor. By the help of some special choices 
(\ref{choice-L}), (\ref{choice-d0}) and (\ref{choice-d1})
we reduced the number
of constants to 3 material constants, namely 2 elastic constants and 
1 internal length. As a result we obtain a dislocation theory with
symmetric force stresses.
The special choice~(\ref{choice-L}) for the three coefficients $a_1$, $a_2$ and $a_3$ 
in the dislocation gauge Lagrangian gives the core energy of the 
edge dislocation in an infinite medium which is in agreement with the 
expression obtained in the Peierls-Nabarro model.
 It is important to note that one can also use the total Lagrangian 
of this dislocation model
%%%%${\cal L}={\cal L}_{\rm disl}+{\cal L}_{\rm grad}+{\cal L}_{\rm grad-disl}
%%%-W+W_{\rm bg}$ 
for the gauge theoretical description of screw dislocations. 
In this turn, one reproduces all quantities of the screw dislocation 
calculated in~\cite{Lazar02}. Therefore, by using the Lagrangian proposed in
this paper, one is able to obtain gauge theoretical solutions for
the edge and screw dislocation which have no singularities and agree 
with the classical solutions in the far field.
The (linear) gauge equation formulated in terms of the stress field is an 
inhomogeneous Helmholtz equation for the screw and the edge dislocation.

It will be interesting to apply  
the nonsingular solutions of the edge and screw dislocation 
to dislocation modelling. One can use the gauge theoretical 
displacement fields to simulate the core region for HRTEM micrographs.
Another interesting application could be the x-ray diffraction profile
analysis.
One should use the materials tungsten, diamond or aluminium 
because they are nearly isotropic. In this way the coefficient $\kappa^{-1}$ should be 
experimentally verified.
Additionally, it would be of interest to compare experimental results 
with the values from different theoretical approaches (gauge theory, strain gradient 
elasticity and/or nonlocal elasticity).
Another possibility to determine the coefficient $\kappa^{-1}$ should be 
a lattice-theoretical approach. The coefficient $\kappa^{-1}$ expresses the interactions and 
the properties in the core which are different from those in the 
undeformed crystal. Consequently, the elastic moduli say nothing about the 
interactions in the core of dislocations.
It should be possible to calculate $\kappa^{-1}$ from the 
coupling parameters of the lattice at the dislocation core. 

\subsection*{Acknowledgement}
The author wishes to thank Professors Friedrich W.~Hehl, Stefan M{\"u}ller 
and Alfred Seeger and Dr. Gerald Wagner for stimulating discussions, 
furthermore to Drs. Mikhail~Yu. Gutkin and Cyril Malyshev for correspondence
and useful remarks.
He would like to express his gratitude to the Max-Planck-Institut f{\"u}r 
Mathematik in den Naturwissenschaften for the fine conditions of work 
and for financial support.


\begin{thebibliography}{99}
\bibitem{Lazar00} M.~Lazar, 
%%%      {\it Dislocation theory as a $3$-dimensional translation gauge theory}, 
        Ann. Phys.~(Leipzig)~{\bf 9} (2000) 461.
%%%%%%% {\tt cond-math$/$0006280}.
\bibitem{Lazar02} M.~Lazar, 
%%%%%    {\it An elastoplastic theory of dislocations as a physical field theory with torsion}, 
        J. Phys. A: Math. Gen.~{\bf 35} (2002) 1983.
%%%%%%% {\tt cond-math$/$0105270}.
\bibitem{Lazar02b} M.~Lazar, 
%%%%    {\it Screw Dislocations in the Field Theory of Elastoplasticity}, 
	Ann. Phys.~(Leipzig)~{\bf 11} (2002) 635.
%%%%%%  {\tt cond-math$/$0203058}.
\bibitem{Malyshev00} C.~Malyshev, 
%%%%    {\it $T(3)$-gauge model, the Einstein-like gauge equation, and Volterra dislocations with modified asymptotics},
         Ann.~Phys. (N.Y.)~{\bf 286} (2000) 249.
\bibitem{GL79} A.A.~Golebiewska-Lasota, 
        Int. J. Engng. Sci.~{\bf 17} (1979) 329.
\bibitem{Edelen82} D.G.B.~Edelen, 
        Int. J. Engng. Sci.~{\bf 20} (1982) 49.
\bibitem{Edelen83} A.~Kadi{\'c} and D.G.B. Edelen, {\it A Gauge Theory of Dislocations and Disclinations},
        in: {\it Lecture Notes in Physics}, Vol. 174, Springer, Berlin (1983).
\bibitem{Edelen88} D.G.B.~Edelen and D.C.~Lagoudas, {\it Gauge Theory and Defects in 
        Solids}, 
        %%%%in {\it Mechanics and Physics of Discrete System}, Vol.~1, G.C.~Sih, ed., 
        North-Holland, Amsterdam (1988).
\bibitem{VS88} M.C.~Valsakumar and D.~Sahoo, Bull. Mater. Sci.~{\bf 10} (1988) 3.
\bibitem{Edelen96} D.G.B.~Edelen, 
        Int. J. Engng. Sci.~{\bf 34} (1996) 81.
\bibitem{Popov} V.L.~Popov, 
        Int. J. Engng. Sci.~{\bf 30} (1992) 329.
\bibitem{Kadic94} A.~Kadi{\'c}-Galeb and R.C.~Batra,
        Int. J. Engng. Sci.~{\bf 32} (1994) 291.
\bibitem{KV92} M.O.~Katanaev and I.V.~Volovich, 
        Ann.~Phys.~(N.Y.)~{\bf 216} (1992) 1.
\bibitem{Gairola81} B.K.D.~Gairola,
        {\it Gauge Invariant Formulation of Continuum Theory of Defects}, in:
        Continuum Models and Discrete Systems, Proc. 4th Int. Symp.,  
        O.~Brulin and R.K.T.~Hsieh, eds., North-Holland, Amsterdam (1981) p.~55.
\bibitem{Gairola93} B.K.D.~Gairola, {\it Gauge Theory of Dislocations}, in:
        Continuum Models and Discrete Systems, Proc. 7th Int. Symp., Paderborn, Germany, 
        K.-H.~Anthony and H.-J.~Wagner, eds.,
        Trans. Techn. Publ., Aedermannsdorf (CH) (1993) p.~579.
\bibitem{Kroener93} E.~Kr{\"o}ner, {\it A Variational Principle in Nonlinear Dislocations
        Theory}, in: Proc. 2nd Int. Conf. Nonlin. Mechanics,  
        Chien Wei-zang, ed., Peking University Press, Beijing (1993) p.~59.
\bibitem{Kroener96} E.~Kr{\"o}ner, {\it Dislocation Theory as a Physical Field Theory}, in:
        Continuum Models and Discrete Systems, Proc. 8th Int. Symp., Varna, Bulgaria, 
        K.Z. Markov, ed., World Scientific, Singapore (1996) p.~522.
\bibitem{Kunin85} I.A.~Kunin, {\it On the Gauge Theory of Dislocations}, in: {\it The Mechanics of Dislocations},
        Eds. E.C.~Aifantis and J.P.~Hirth, American Society of Metals, 
        Metals Park, Ohio (1985) p.~69.
\bibitem{Kleinert89} H.~Kleinert, {\it Gauge Fields in Condensed Matter Vol. II:
        Stresses and Defects}, World Scientific, Singapore (1989). 
\bibitem{Osipov} V.A.~Osipov, J. Phys. A: Math. Gen.~{\bf 24} (1991) 3237.
\bibitem{Volkov} N.B.~Volkov, J. Phys. A: Math. Gen.~{\bf 30} (1997) 6391.
\bibitem{Bogatov} N.M.~Bogatov, Phys. Stat. Sol.~(b)~{\bf 228} (2001) 651.
\bibitem{Peierls40} R.E.~Peierls, Proc. Phys. Soc. (London)~{\bf 54} (1940) 34.
\bibitem{Nabarro47} F.R.N.~Nabarro, Proc. Phys. Soc. (London)~{\bf 59} (1947) 256.
\bibitem{Nabarro} F.R.N.~Nabarro, {\it Theory of Crystal Dislocations}, Dover,
        New York (1987).
\bibitem{HL} J.P.~Hirth and J.~Lothe, {\it Theory of Dislocations}, %%%%%2nd edition, 
	John Wiley, New York (1982).
\bibitem{Eringen77} A.C.~Eringen, Int. J. Engng. Sci.~{\bf 15} (1977) 177.
\bibitem{Eringen83} A.C.~Eringen, J. Appl. Phys.~{\bf 54} (1983) 4703.
\bibitem{AE83} N.~Ari and A.C.~Eringen, Cryst. Lattice Defects Amorph. Mat.~{\bf 10} 
        (1983) 33.
\bibitem{Eringen85} A.C.~Eringen, {\it Nonlocal Continuum Theory for Dislocations
        and Fracture}, in: {\it The Mechanics of Dislocations},
        Eds. E.C.~Aifantis and J.P.~Hirth, American Society of Metals, 
        Metals Park, Ohio (1985) p.~101.
\bibitem{Eringen87} A.C.~Eringen, 
%%%%%   {\it Theory of nonlocal elasticity and some applications}, 
        Res. Mech.~{\bf 21} (1987) 313.
%%%\bibitem{Eringen90} A.C.~Eringen, {\it On Screw Dislocations and Yield}, in:
%%%     Elasticity, Mathematical Methods and Applications, G.G.~Eason and 
%%%     R.W.~Odgen, eds., Ellis Harwood, Chichester (1990) p.~87.
\bibitem{AA92} B.S.~Altan and E.C.~Aifantis, Scripta Metall. Mater.~{\bf 26} 
        (1992) 319.
\bibitem{Aifantis94} E.C.~Aifantis, J.~Mechan. Behav. Mater.~{\bf 5} 
        (1994) 355.
\bibitem{AA97} B.S.~Altan and E.C.~Aifantis, J.~Mech. Behav. Mater.~{\bf 8} 
        (1997) 231.
\bibitem{GA96} M.Yu.~Gutkin and E.C.~Aifantis, 
%%%%%   {\it Screw dislocation in gradient elasticity}, 
        Scripta Mater.~{\bf 35} (1996) 1353.
\bibitem{GA97} M.Yu.~Gutkin and E.C.~Aifantis, 
%%%%    {\it Edge dislocation in gradient elasticity}, 
        Scripta Mater.~{\bf 36} (1997) 129.
\bibitem{GA99} M.Yu.~Gutkin and E.C.~Aifantis, 
%%%%    {\it Dislocations in the theory of gradient elasticity},
        Scripta Mater.~{\bf 40} (1999) 559.
%%%%\bibitem{GA99b} M.Yu.~Gutkin and E.C.~Aifantis, 
%%%%    {\it Dislocations and disclinations in gradient elasticity},
%%%%    Phys. Stat. Sol. (b)~{\bf 214} (1999) 245.
\bibitem{Gutkin00} M.Yu.~Gutkin, Rev. Adv. Mater. Sci.~{\bf 1} (2000) 27.
%%%\bibitem{Fleck94} N.A.~Fleck, G.M.~Muller, M.F.~Ashby and J.W.~Hutchinson,
%%%        Acta Met. et Mat.~{\bf 42} (1994) 475.
\bibitem{Brailsford66} A.D.~Brailsford, Phys. Rev.~{\bf 142} (1966) 383.
\bibitem{Kunin86} I.A.~Kunin, {\it Theory of Elastic Media with Microstructure},
        Springer, Berlin (1986).
\bibitem{VK93} G.~V{\"o}r{\"o}s and I.~Kov{\'a}cs, 
        Phys. Stat. Sol.~(b)~{\bf 178} (1993) 99.
\bibitem{VK95} G.~V{\"o}r{\"o}s and I.~Kov{\'a}cs, 
        Phil. Mag.~A~{\bf 72} (1995) 949.
\bibitem{Gunter58} W.~G{\"u}nther, {\it Zur Statik und Kinematik des Cosseratschen 
	Kontinuums}, Abh. Braunschweig. Wiss. Ges., Bd.~{\bf 10} (1958) 195.
\bibitem{Schaefer67} H.~Schaefer, Z.~Ang.~Math.~Mech.~{\bf 47} (1967) 485.
\bibitem{Kluge69} G.~Kluge, Wiss. Z. Techn. Hoch.~Magdeburg~{\bf 13} (1969) 377.
\bibitem{Misicu65} M.~Mi\c{s}icu, Rev.~Roum.~Sci.~Techn., S\'{e}r.~m\'{e}c.~appl.~{\bf 10} 
	(1965) 35.
\bibitem{Teodosiu65} C.~Teodosiu, Rev.~Roum.~Sci.~Techn., S\'{e}r.~m\'{e}c.~appl.~{\bf 10} 
	(1965) 1461.
\bibitem{Kessel70} S.~Kessel, Z.~Ang.~Math.~Mech.~{\bf 50} (1970) 547.
\bibitem{Knesl72} Z.~Kn\'{e}sl and F.~Semela, Int.~J.~Eng.~Sci.~{\bf 10} (1972) 83.
\bibitem{Nowacki73} J.P.~Nowacki, Bull. Acad. Polon. Sci., S{\'e}r. sci. techn.~{\bf 21}
        (1973) 585.
\bibitem{Nowacki74} N.~Nowacki, Arch. Mech.~{\bf 26} (1974) 3.
\bibitem{Minagawa77} S.~Minagawa, Appl.~Eng.~Sci.~Lett.~{\bf 5} (1977) 85.
\bibitem{Edelen89} D.G.B.~Edelen, 
        Int. J. Engng. Sci.~{\bf 27} (1989) 653.
\bibitem{Kondo52} K.~Kondo, {\it On the Geometrical and Physical Foundations of the Theory
        of Yielding}, in {\it Proceedings of the 2nd Japan National Congress for Applied
        Mechanics}, Tokyo (1952) p.~41.
\bibitem{Bilby55} B.A.~Bilby, R.~Bullough and E.~Smith,
%%%%      {\it Continous distributions of dislocations: 
%%%%    A new application of the method of non-Riemannian geometry}, 
        Proc. Roy. Soc.~(London)~A~{\bf 231} (1955) 263.
\bibitem{KS59} E.~Kr{\"o}ner and A.~Seeger,
%%%%     {\it Nicht-lineare Elastizit{\"a}tstheorie der Versetzungen und Eigenspannungen}, 
        Arch. Rat. Mech. Anal.~{\bf 3} (1959) 97.
\bibitem{Kroener60} E.~Kr{\"o}ner, 
%%%%    {\it Allgemeine Kontinuumstheorie der Versetzungen und Eigenspannungen}, 
        Arch. Rat. Mech. Anal.~{\bf 4} (1960) 273. 
\bibitem{MAG} F.W.~Hehl, J.D. McCrea, E.W. Mielke and Y. Ne'eman,
%%%%    {\it Metric--affine gauge theory of gravity: Field equations, Noether
%%%%    identities, world spinors, and breaking of dilation invariance},
        Phys. Rep.~{\bf 258} (1995)~1.
\bibitem{Nye} J.F.~Nye, Acta Met.~{\bf 1} (1953) 153.
\bibitem{deWit70} R.~deWit, {\it Linear Theory of Static Disclinations}, 
        in {\it Fundamental Aspects of Dislocation Theory}, 
        Vol.~1, Eds. J.A.~Simmons, R.~deWit and R.~Bullough, Nat. Bur. Stand. (U.S.), 
        Spec. Publ.~{\bf 317} (1970) p.~651.
\bibitem{deWit73} R.~deWit, J.~Res.~Nat. Bur. Standards~{\bf 77A} (1973) 49.
%%%\bibitem{Kroener69} E.~Kr{\"o}ner, {\it Plastizit{\"a}t und Versetzungen}, in:
%%%	{\it Mechanik der deformierten Medien} by A.~Sommerfeld, Harri
%%%	Deutsch, Frankfurt/M. (1992) reprint of the 6th ed. (1969) pp.~310-376. 
\bibitem{Mindlin65} R.D.~Mindlin, Int. J. Solids Struct.~{\bf 1} (1965) 265.
\bibitem{Maugin} G.A.~Maugin, {\it Material Inhomogeneities in Elasticity}, 
	Chapman and Hall, London (1993).
\bibitem{Toupin64} R.A.~Toupin, Arch. Rat. Mech. Anal.~{\bf 17} (1964) 85. 
\bibitem{BA56} W.L.~Bond and J.~Andrus, Phys. Rev.~{\bf 101} (1956) 1211.
\bibitem{Bullough58} R.~Bullough, Phys. Rev.~{\bf 110} (1958) 620.
\bibitem{ND70} V.I.~Nikitenko and L.M.~Dedukh, Phys. Stat. Sol.~(a)~{\bf 3} 
        (1970) 383.
\bibitem{Timpe05} A.~Timpe, Zeits. f. Math. Phys.~{\bf 52} (1905) 348.
\bibitem{Brown41} W.F.~Brown, Phys. Rev.~{\bf 60} (1941) 139.
\bibitem{Koehler41} J.S.~Koehler, Phys. Rev.~{\bf 60} (1941) 397.
\bibitem{LL49} G.~Leibfried and K.~L{\"u}cke, Z.~Phys.~{\bf 126} (1949) 450.
\bibitem{Kroener58} E.~Kr{\"o}ner, {\it Kontinuumstheorie der Versetzungen und Eigenspannungen},
        Erg.~Angew. Math.~{\bf 5} (1958) p.~1.
\bibitem{Kroener81} E.~Kr{\"o}ner, {\it Continuum Theory of Defects}, in:
        {\it Physics of Defects} (Les Houches, Session 35), R.~Balian et al., eds.,
        North-Holland, Amsterdam (1981) p.~215.
\bibitem{deWit73b} R.~deWit, J.~Res.~Nat. Bur. Standards~{\bf 77A} (1973) 607.
\bibitem{Read} W.T.~Read, {\it Dislocations in Crystals}, McGraw Hill, 
        New York (1953).
\bibitem{Teodosiu} C.~Teodosiu, {\it Elastic Models of Crystal Defects},
        Springer-Verlag, Berlin (1982).
\bibitem{HK65} F.W.~Hehl and E.~Kr{\"o}ner, Z.~Naturforschg.~{\bf 20a} (1965) 336.
%%%\bibitem{Miller98} R.~Miller, R.~Phillips, G.~Beltz and M.~Ortiz, 
%%%        J.~Mech. Phys. Sol.~{\bf 46} (1998) 1845.
\bibitem{Kroener63} E.~Kr{\"o}ner, Int. J. Engng. Sci.~{\bf 1} (1963) 261.
\bibitem{Kroener66} E.~Kr{\"o}ner and B.K.~Datta, Z. Phys.~{\bf 196} (1966) 203.
\bibitem{Kroener67} E.~Kr{\"o}ner, Int. J. Solids Struct.~{\bf 3} (1967) 731.
\end{thebibliography}
\end{document}